\documentclass[fleqn,usenatbib]{mnras}

\usepackage[T1]{fontenc}

\usepackage{graphicx}	% Including figure files
\usepackage{amsmath}	% Advanced maths commands
\usepackage{multirow}
\usepackage{pdflscape}
\usepackage{rotfloat}
\usepackage{comment}
\usepackage{supertabular}
\usepackage{longtable}

\title[The first carbon isotopic ratios in post-RGB stars]{The first measurements of carbon isotopic ratios in post-RGB stars: SZ Mon and DF Cyg\\
{\small E-iSpec: A spectral analysis tool to derive elemental abundances and isotopic ratios for evolved stars}} %http://mirrors.ibiblio.org/CTAN/macros/latex/contrib/mnras/mnras_guide.pdf

\author[Mohorian et al. (2024)]{Maksym Mohorian,$^{1,2}$\thanks{E-mail: maksym.mohorian@hdr.mq.edu.au} Devika Kamath,$^{1,2,3}$ Meghna Menon,$^{1,2}$ Paolo Ventura,$^{3,4}$  \newauthor
Hans Van Winckel$^{5}$, D. A. Garc\'{i}a-Hern\'{a}ndez$^{6,7}$, and Thomas Masseron$^{6,7}$
\\
$^{1}$School of Mathematical and Physical Sciences, Macquarie University, Balaclava Road, Sydney, NSW 2109, Australia\\
$^{2}$Astrophysics and Space Technologies Research Centre, Macquarie University, Balaclava Road, Sydney, NSW 2109, Australia\\
$^{3}$INAF, Osservatorio Astronomico di Roma, Via Frascati 33, I-00077 Monte Porzio Catone, Italy\\
$^{4}$Osservatorio Astronomico di Roma: Monte Porzio Catone, Lazio, Italy\\
$^{5}$Institute of Astronomy, KU Leuven, Celestijnenlaan 200D, 3001 Leuven, Belgium\\
$^{6}$Instituto de Astrof\'{i}sica de Canarias (IAC), E-38205 La Laguna, Tenerife, Spain\\
$^{7}$Departamento de Astrof\'{i}sica, Universidad de La Laguna (ULL), E-38206 La Laguna, Tenerife, Spain
}

\date{Accepted XXX. Received YYY; in original form ZZZ}

\pubyear{2024}

\begin{document}\label{firstpage}
\pagerange{\pageref{firstpage}--\pageref{lastpage}}
\maketitle
\begin{abstract}
Dusty post-red giant branch (post-RGB) stars are low- and intermediate-mass stars where the RGB evolution was prematurely terminated by a poorly understood binary interaction. These binary stars are considered to be low-luminosity analogues of post-asymptotic giant branch (post-AGB) binary stars. In this study, we investigated the chemical composition of two dusty post-RGB binary stars, SZ~Mon and DF~Cyg, using multi-wavelength spectroscopic data from HERMES/Mercator (optical) and the APOGEE survey (near-infrared). Owing to challenges posed by existing spectral analysis tools for the study of evolved stars with complex atmospheres, we developed E-iSpec: a dedicated spectral analysis tool for evolved stars, to consistently determine atmospheric parameters, elemental abundances, and carbon isotopic ratios. Our abundance analysis revealed that observed depletion patterns and estimated depletion efficiencies resemble those found in post-AGB binary stars. However, the onset of chemical depletion in post-RGB targets occurs at higher condensation temperatures ($T_{\rm turn-off,~post-RGB}\approx1\,400$~K), than in most post-AGB stars ($T_{\rm turn-off,~post-AGB}\approx1\,100$~K). Additionally, our study resulted in the first estimates of carbon isotopic ratios for post-RGB stars ($^{12}$C/$^{13}$C$_{\rm SZ~Mon}=8\pm4$, $^{12}$C/$^{13}$C$_{\rm DF~Cyg}=12\pm3$). We found that the observationally derived CNO abundances and the carbon isotopic ratios of our post-RGB binary targets are in good agreement with theoretical predictions from the ATON single star evolutionary models involving first dredge-up and moderately-deep extra mixing. This agreement emphasises that in post-RGB binary targets, the observed CNO abundances reflect the chemical composition expected from single star nucleosynthesis (i.e., convective and non-convective mixing processes) occurring during the RGB phase before it is terminated.
\end{abstract}

\begin{keywords}
\textit{stars: evolution, stars: binaries, stars: AGB and post-AGB, stars: chemically peculiar, stars: abundances, techniques: spectroscopic}
\end{keywords}

\section{Introduction}\label{sec:int}
Low- and intermediate-mass (LIM) stars, which have masses roughly between 0.9 to 8~M$_\odot$, account for $\sim95\%$ of the aging stellar population and produce $\sim90\%$ of the ejected material in terms of silicates and carbonaceous dust \citep{sloan2008dust}. This makes LIM stars key contributors to the chemical enrichment of the Universe \citep{kobayashi2020OriginOfElements}.

For single LIM stars, during their giant branch phases, i.e., the red giant branch (RGB) phase and the asymptotic giant branch (AGB) phase, the chemical elements synthesised through nuclear fusion in stellar interiors are brought to the surface through convective (and non-convective) mixing processes \citep{busso2007mixing, karakas2014dawes, ventura2015HBB, ventura2018mixing}. However, the physical mechanisms that govern the  nucleosynthesis of elements and elemental isotopes, different mixing processes and mass loss, remain poorly understood \citep{ventura2022InternalProcesses}.

Based on past theoretical studies \citep[see][and references therein]{karakas2014dawes}, it is widely accepted that the first significant alteration in the photospheric chemical composition of LIM stars occurs during the RGB phase, primarily due to convective-driven mixing processes, commonly referred to as first dredge-ups (FDUs). The observational studies of single RGB stars \citep{carbon1982FDU, pilachowski1986FDU, kraft1994FDU, shetrone2019FDU} further evidenced that the FDU leads to an enrichment of surface abundances of $^4$He, $^{13}$C, $^{14}$N, $^{17}$O, and $^{23}$Na, while causing a depletion in surface abundances of $^7$Li, $^{12}$C, $^{16}$O, and $^{18}$O. Overall, the mixing processes on RGB lead to a substantial decrease in the $^{12}$C/$^{13}$C, $^{14}$N/$^{15}$N, and $^{16}$O/$^{17}$O isotopic ratios and slight increase in $^{16}$O/$^{18}$O isotopic ratio \citep{abia2017CNOisotopesAGB, mccormick2023RGBExtraMixing}. Moreover, subsequent observational studies have revealed peculiar abundance patterns of elements, such as He, Li, C, N, O, and Na, on the RGB, where the FDU alone fails to account for the observed changes in these elements \citep{sneden1986ExtraMixing, gilroy1989ExtraMixing, smiljanic2009ExtraMixing, tautvaisiene2013ExtraMixing, drazdauskas2016ExtraMixing, charbonnel2020ExtraMixing}. To address the observationally derived abundances from RGB stars, theoretical models have incorporated non-convective `extra' mixing processes, such as rotation-induced mixing, thermohaline mixing, meridional circulation, shear mixing, and various hydrodynamic and magnetic mixing processes \citep[e.g.,][and references therein]{palacios2003ExMixMod, charbonnel2007ExMixMod, lagarde2012ExMixMod, karakas2014dawes}.

In this work, we aim to trace the evolutionary processes and nucleosynthesis that occur during the RGB phase using dusty post-red giant branch (post-RGB) stars \citep{kamath2016postRGBs}. Dusty post-RGB stars are low-luminosity \citep[with luminosities below the tip of the RGB, $L_{\rm RGB\ tip}\approx2\,500 L_\odot$;][]{2014MNRAS.439.2211K, kamath2022GalacticSingles} analogues of binary post-asymptotic giant branch (post-AGB) stars, where their RGB phase was likely terminated by binary interaction. Post-RGB stars therefore occupy a lower luminosity region in the Hertzsprung-Russell (HR) diagram (compared to post-AGB stars), situated bluewards of the RGB, with the presence of circumstellar dust attributed to the mass loss process induced by binary interaction \cite{kamath2019depletionLMC}.

Previously, several studies \citep[e.g.,][]{desmedt2012j004441, desmedt2015lmc2obj, kamath2019depletionLMC, dellagli2023silicates} extensively explored low-mass stars in the post-AGB phase as valuable tracers of evolution, nucleosynthesis, mass loss through thermal pulses, and mixing processes that occur during the AGB phase. More recently, the study by \cite{kamath2023models} showed that the luminosity and the surface carbon abundance of a post-AGB star serve as the most valuable indicators, unveiling the preceding evolution and nucleosynthetic history of the star. \cite{kamath2023models} derived the masses of AGB progenitors for 31 single post-AGB stars by comparing their observed chemical composition with the predictions from ATON stellar evolutionary models. This highlights the wealth of using post-AGB stars as tracers of AGB nucleosynthesis. In this study, we extend the concept of using post-AGB stars as tracers of nucleosynthesis by leveraging observations of their low-luminosity analogues -- dusty post-RGB stars.
 
The dusty post-RGB stars display properties resembling post-AGB binary stars. For example, they have similar atmospheric parameters such as effective temperatures ($T_{\rm eff}$) ranging from 3\,500 to 8\,500 K, surface gravities ($\log g$) ranging from 0 to 2.5 dex, and metallicities ranging from --5 to 0 dex (depending on their host galaxy). Comparable to post-AGB binary stars, the majority of the post-RGB stars exhibit a `disc-type' spectral energy distribution (SED), characterized by a near-infrared (NIR) excess, indicative of the presence of a circumbinary disc \citep[][and references therein]{kluska2022GalacticBinaries}. Furthermore, a subset of post-RGB stars also demonstrates Type II Cepheid variability \citep{2014MNRAS.439.2211K, 2015MNRAS.454.1468K, kamath2016postRGBs, manick2019RVTauDFCyg}.

Post-RGB stars and binary post-AGB stars also share a common characteristic in their photospheric chemistry known as `chemical depletion'. This chemical anomaly was first identified through various studies focusing on post-AGB stars \citep[such as][]{waelkens1991depletion, vanwinckel1992depletion, venn2014depletion, kamath2019depletionLMC, oomen2019depletion, kluska2022GalacticBinaries}, and then confirmed for post-RGB stars by piecemeal studies \citep[e.g.,][]{giridhar2005rvtau, maas2007t2cep, manick2019RVTauDFCyg, gezer2019depletion}.

Chemical depletion is characterised by a photospheric abundance pattern resembling the patterns observed in the interstellar gas phase \citep[][and references therein]{konstantopoulou2022ISMdepletion}. In this pattern, refractory elements like Al, Ti, Fe, Y, and La are typically underabundant, whereas volatile elements such as Zn and S tend to maintain their initial abundance levels, similar to those observed in the interstellar medium. Consequently, these stars exhibit extrinsically metal-poor characteristics with [Fe/H] ranging from --5.0 to --0.5 \citep{vanwinckel1992depletion, oomen2019depletion}. However, it is important to note that the abundances of C, N, and O deviate from the pattern of other volatile elements. This deviation arises from likely alterations in CNO abundances due to dredge-ups occurring during previous RGB and AGB evolutionary phases.

In this study, we targeted two post-RGB stars that are known to show photospheric chemical depletion \citep{maas2007t2cep, giridhar2005rvtau}. We used high-resolution optical and mid-resolution NIR spectra to derive accurate elemental abundances and carbon isotopic ratios. Our goal is to investigate whether the combined knowledge of the luminosity together with the surface carbon and nitrogen (i.e., CNO abundances, the C/O and $^{12}$C/$^{13}$C ratios) could serve as valuable tracers of the nucleosynthetic history that occurs during the RGB phase before transitioning to the post-RGB phase. This study also let us investigate whether binary interactions cause any drastic effect on the mixing processes that are known to alter the surface chemical composition.

In Section~\ref{sec:tar}, we provide an overview of the targets under investigation. In Section~\ref{sec:obs}, we introduce the data and observations. In Section~\ref{sec:san}, we describe E-iSpec, a dedicated spectral analysis tool developed for this study to derive elemental abundances and isotopic ratios in evolved stars, and present our research findings. In Section~\ref{sec:mod}, we compare observationally derived abundances and carbon isotopic ratios with theoretical predictions from the ATON evolutionary models \citep{ventura2008aton3} of single RGB stars that have similar atmospheric parameters and luminosities as our binary post-RGB targets. Finally, in Section~\ref{sec:con}, we delve into the potential of using observationally derived atmospheric parameters, elemental abundances, and isotopic ratios to explore the impact of binary interactions on RGB nucleosynthesis and mixing processes.

\section{Target sample}\label{sec:tar}
The target sample for this study comprised of two binary post-RGB stars: SZ~Mon and DF~Cyg (see Table~\ref{tab:varpro}), which were derived from a larger sample of spectroscopically verified evolved candidates in the Galaxy and the Magellanic Clouds (see Appendix~\ref{app:tar} for more details on the target selection). We note that both targets are Type II Cepheid variables \citep{maas2007t2cep, giridhar2005rvtau}.

We present target details individually in the following subsections.

\begin{table}
    \centering
    \caption{Variability properties and luminosity estimates (see Section~\ref{ssec:obslum}) obtained in this study for our post-RGB binary pulsating variables, SZ Mon and DF Cyg.\\} \label{tab:varpro}
    \begin{tabular}{lcc}\hline
         Parameter & SZ Mon & DF Cyg \\\hline
         RVb phenomenon & no$^{a}$ & yes$^{b}$ \\
         $P_\textrm{puls}$ (days) & 16.34$^{c}$ & 24.91$^{d}$ \\
         $P_\textrm{orb}$ (days) & -- & 780$^{d}$\\\hline
         L$_{\rm SED}$ (L$_\odot$) & $193\pm30$ & $657\pm103$ \\
         L$_{\rm PLC}$ (L$_\odot$) & $382\pm83$ & -- \\\hline
    \end{tabular}\\
    %\raggedright
    \textbf{Notes:} $P_\textrm{puls}$ is the fundamental pulsation period, $P_\textrm{orb}$ is the orbital period, L$_{\rm SED}$ is the luminosity derived from the SED fitting, L$_{\rm PLC}$ is the luminosity calculated from the PLC relation. The periods and the detection of RVb phenomenon are acquired from the following catalogues: $^a$\cite{kluska2022GalacticBinaries}, $^b$\cite{manick2019RVTauDFCyg}, $^c$\cite{pawlak2019ASASDFCyg}, $^d$\cite{kissbodi2017RVTau}.\\
\end{table}

\subsection{SZ Mon}
SZ Mon is a spectral type F8 star showing broad infrared excess in the SED. The very first studies of SZ~Mon \citep{lloyd1968szmon, stobie1970szmon} classified this object as a Type II Cepheid with a double period\footnote{Double period is the empirically best-fit value for phase curves of Type II Cepheids, yet the fundamental pulsation period is twice shorter \citep{stobie1970szmon}.} of 32.686 days. \cite{maas2007t2cep} classified SZ~Mon as RVa photometric type, i.e., an RV Tau variable without a detected secondary variation of its phase curve with significant amplitude and long period $\sim1000$d (referred to as the RVb phenomenon).

In the recent studies \citep{maas2007t2cep, oomen2019depletion}, SZ Mon was classified as a post-AGB star with a disc-type SED. \cite{oomen2019depletion} presented the SED modelling for SZ~Mon and derived the SED luminosity $L_\textrm{SZ Mon, SED} = 2400^{+2000}_{-1000}L_\odot$.

Furthermore, chemical abundance studies by \cite{maas2007t2cep} showed that SZ~Mon is an evolved post-AGB binary star showing photospheric chemical depletion -- characteristic of post-AGB and post-RGB binary stars (see Section~\ref{sec:int}).

We note that, while SZ~Mon was classified as a post-AGB RV Tauri (RV Tau) variable in the literature, in this study we re-classified the evolutionary nature of this target. Based on the newly derived luminosity of SZ~Mon (see Section~\ref{ssec:obslum}), its disc-type SED, and the confirmation of photospheric chemical depletion (see Section~\ref{ssec:sanabs}), we conclude that SZ~Mon is a post-RGB binary star. Furthermore, given that the observed fundamental pulsation period is less than 20 days \citep[i.e., 16.336 days;][]{soszynski2008OGLE}, SZ~Mon rather belongs to the W Virginis (W Vir) subclass of Type II Cepheid family (instead of RV Tau subclass). We adopted this new classification of SZ~Mon (i.e., a post-RGB binary star and W Vir variable) for the rest of this study.

\subsection{DF Cyg}
DF Cyg is a star of G6/7Ib/IIa spectral type. DF~Cyg exhibits a broad infrared excess in the SED \citep[i.e., disc-type][]{deruyter2006discs}. \cite{manick2019RVTauDFCyg} derived the luminosity of DF~Cyg from SED fitting, wherein $L_\textrm{DF Cyg, SED} = 1010^{+150}_{-140}L_\odot$).

\cite{giridhar2005rvtau} conducted an extensive abundance analysis of DF~Cyg and identified a rather steep depletion pattern with an unusually negative volatile-refractory abundance ratio ([Zn/Ti]=$-0.7$ dex). In our current study (see Section~\ref{ssec:sanabs}), we performed an independent abundance analysis, which resulted in a standard depletion pattern with ([Zn/Ti]=$+0.35$ dex).

DF~Cyg was classified as an RV Tau variable with a fundamental pulsation period of 24.91 days \citep{kissbodi2017RVTau}. Using the pulsation period, \cite{manick2019RVTauDFCyg} derived a more reliable estimate of DF~Cyg luminosity from period-luminosity-colour (PLC) relation for Type II Cepheids $L_\textrm{DF Cyg, PLC} = 990\pm190L_\odot$), and classified DF~Cyg as a post-RGB star.

Previous studies have suggested that DF~Cyg could be a binary system due to the very strong RVb phenomenon in the light curve, as discussed by \cite{bodi2016dfcyg} and \cite{vega2017dfcyg}, or due to the presence of a disc, as mentioned by \cite{deruyter2005discs} and explained in \cite{winckel2018}. Orbital parameter analysis \citep{oomen2018OrbitalParameters} confirmed the binary nature of DF~Cyg with an orbital period of 784 days (see Table~\ref{tab:varpro}).

In this paper, we validated the binary post-RGB nature of DF~Cyg (see Section~\ref{ssec:obslum}).

\section{Data and observations}\label{sec:obs}
In this section, we present the photometric data used for SED fitting (see Section~\ref{ssec:obspht}) and luminosity derivation (see Section~\ref{ssec:obslum}), as well as the spectroscopic data used for deriving elemental abundances and isotopic ratios (see Section~\ref{ssec:obsspc}).

\subsection{Photometric data}\label{ssec:obspht}
In Table~\ref{tab:phomag}, we present the photometric magnitudes of SZ~Mon and DF~Cyg in various wavelength bands, including optical, NIR, and mid-infrared (MIR). The UBVRI photometry is taken from the Johnson-Cousins system \citep{johnson1953,cousins1976}, and the I-band filters from the Sloan Digital Sky Survey \citep[SDSS, ][]{SDSS2000photometry}. The Two Micron All Sky Survey (2MASS) provides magnitudes in the J, H, and K bands \citep[1.24, 1.66, and 2.16 $\mu$m, respectively; ][]{2MASS2006}. WISE \citep{WISE2010} contains the magnitudes in the W1, W2, W3, and W4 bands (3.4, 4.6, 12, and 22 $\mu$m, respectively). The mid-infrared (MIR) was complemented by fluxes from the MSX, AKARI, and IRAS catalogues \citep{MSX2003,AKARI2010,IRAS1984}.% To compensate for foreground extinction, we assume that the wavelength dependency of the total extinction in the line of sight follows the interstellar-medium extinction law with R$_{\rm V}$=3.1 \cite{SEDextinction1989}.

\begin{figure}
    \centering
    \includegraphics[width=.99\linewidth]{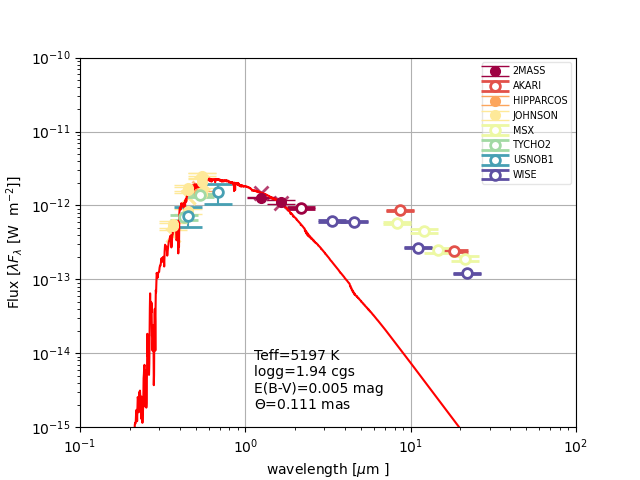}
    \includegraphics[width=.99\linewidth]{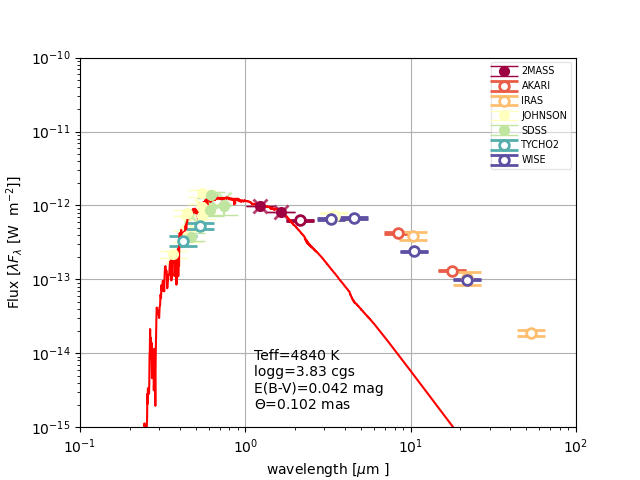}
    \caption{Spectral energy distribution of SZ Mon (top) and DF Cyg (bottom) suggests the possible presence of circumbinary discs in these targets (see Section~\ref{ssec:obspht}). Red fit is the appropriate Kurucz model atmosphere. We note that the photometric observations of both targets were obtained at slightly different pulsation phases. The legend for the symbols and colours used are included within the plot.}\label{fig:SEDs}
\end{figure}

The SEDs of SZ~Mon and DF~Cyg are shown in Fig.~\ref{fig:SEDs} (top and bottom, respectively). We note that in the case of DF~Cyg, we specifically excluded the $V$ and $I$ band magnitudes (which come from the ASAS Catalogue of Variable Stars in the Kepler Field), since they were obtained during the minima in the RVb period, which is out of phase with all the other photometric data points (see Fig. 8 of \cite{manick2018PLC}. This significantly improves the $\chi^2$ value of the SED fitting and provides a more accurate luminosity estimate (see Section~\ref{ssec:obslum}).

The SEDs of SZ~Mon and DF~Cyg exhibit characteristics of a disc type \citep{oomen2018OrbitalParameters}, suggesting the possible presence of a circumbinary disc, which is indicative of binarity in post-RGB and post-AGB stars \citep{kamath2016postRGBs, kluska2022GalacticBinaries}.

\begin{table*}
    \centering
    \caption{Photometric data for SZ Mon and DF Cyg (see Section \ref{ssec:obspht}). For each filter we provide the units and the central wavelengths in $\mu$m. This table is published in its entirety in the electronic edition of the paper. A portion is shown here for guidance regarding its form and content.}\label{tab:phomag}
    \begin{tabular}{|l|c|c|c|c|c|c|c|}\hline
        Filter & JOHNSON.U & JOHNSON.B & ... & AKARI.L18W & MSX.E & WISE.W4 & IRAS.25um \\
        & (mag) & (mag) && (Jy) & (Jy) & (mag) & (Jy) \\
        & 0.364 & 0.443 & ... & 18.9 & 21.5 & 22.1 & 25.0 \\ \hline %& 0.36373 & 0.443354 & ... & 18.9186 & 21.5008 & 22.0789 & $\approx$25 \\ \hline
        SZ Mon & 11.1$\pm$0.1 & 11.33$\pm$0.08 & ... & 1.59$\pm$0.04 & 1.4$\pm$0.1 & 2.41$\pm$0.02 & -- \\
        DF Cyg & -- & 12.8$\pm$0.2 & ... & 0.88$\pm$0.02 & -- & 2.63$\pm$0.02 & 0.8$\pm$0.2 \\ \hline
    \end{tabular}
\end{table*}

\subsection{Determination of luminosities from SED fitting and PLC relation}\label{ssec:obslum}
In this study, we calculated luminosities of SZ~Mon and DF~Cyg using two methods: i) by fitting the SED (SED luminosity, $L_{\rm SED}$); and ii) by using the PLC relation (PLC luminosity, $L_{\rm PLC}$). We briefly describe these two methods below. 

To compute the SED luminosities, we followed the procedure described in \cite{oomen2018OrbitalParameters, kluska2022GalacticBinaries}. In short, we started by assuming the appropriate Kurucz model atmospheres \citep{2003IAUS..210P.A20C} to fit the initial photometric data points. Following this, we accurately computed a dereddened SED model for SZ~Mon and DF~Cyg. This model took into account the overall reddening or extinction parameter $E(B-V)$ that yielded the lowest $\chi^2$ value from our extensive parameter grid search. The total reddening considered here encompassed both interstellar and circumstellar reddening. We made the assumption that the total reddening along the line of sight follows the wavelength dependence described by the interstellar-medium extinction law \citep{cardelli1989extinction} with an $R_V$=3.1. We incorporated the more accurate Bailer-Jones geometric distances, denoted as $z_{\rm BJ}$, along with their respective upper and lower limits, represented as $z_{\rm BJU}$ and $z_{\rm BJL}$, sourced from the research conducted by \cite{bailerjones2021distances}. These geometric distances were computed based on Gaia EDR3 parallaxes, taking into consideration a direction-dependent prior distance. It is important to note that in all our calculations, we assumed that the emission of flux from the stars followed an isotropic radiation pattern. We also note that stellar variability was not considered, and this omission resulted in an increased $\chi^2$ value for high-amplitude variables.

To calculate the PLC luminosity, we used the PLC relation, which was obtained following the similar procedure as in \cite{manick2018PLC}, but updated with OGLE-IV data for RV Tau pulsating variables in the LMC (Menon et al., submitted). In short, this relation uses the colour-corrected V-band magnitude known as the Wesenheit index \citep[${\rm WI}=V-2.55(V-I)_0$; see][]{ngeow2005wesenheit} and is given as
\begin{equation}
    M_{bol, {\rm WI}} = m\times\log P_0 + c - \mu + BC + 2.55\times(V-I)_0,
\end{equation}
where $M_{bol, {\rm WI}}$ is the absolute Wesenheit magnitude, $P_0$ is the observed fundamental pulsation period in days, $\mu=18.49$ is the distance modulus for the LMC, $(V-I)_0$ is the intrinsic colour of each star. $m=-3.59$ is the slope, and $c=18.79$ is the intercept of the linear fit of the relation (see Fig. 5 in \cite{manick2018PLC}). Bolometric corrections $BC$ for SZ\,Mon and DF\,Cyg were calculated based on their observed effective temperatures, $T_{\rm eff}$ = 5\,460 K for SZ\,Mon (see Table~\ref{tab:fnlabsSZM}) and $T_{\rm eff}$ = 5\,770 K for DF\,Cyg (see Table~\ref{tab:fnlabsDFC}), following the tabulated relation from \cite{flower1996BoloCorr}, corrected by \cite{torres2010BoloCorrErrata}.

We would like to highlight that due to the exclusion of the only available I-band magnitude (as discussed in Section~\ref{ssec:obspht}), the calculation of the PLC luminosity for DF~Cyg is not feasible. Nevertheless, our refined photometric selection for the SED fit provides a more reliable luminosity in comparison to the SED and PLC luminosity values reported by \cite{manick2018PLC}.

In Table~\ref{tab:varpro}, we provide the final values and corresponding uncertainties in the SED and the PLC luminosities for SZ~Mon and the SED luminosity for DF~Cyg. The uncertainties in the SED luminosities were determined by calculating the standard deviation of the upper and lower luminosity limits, taking into consideration the uncertainties associated with distances ($z_{\rm BJU}$ and $z_{\rm BJL}$) and reddening ($E(B-V)$) for each respective target. On the contrary, the uncertainties in the PLC luminosities were dominated by the uncertainties of the reddening.

We note that we considered the PLC luminosity to be more precise and more reliable than the SED luminosity, mainly due to the significant $T_{\rm eff}$ fluctuations in our targets throughout the pulsation cycle (leading to significant changes in model fitting of stellar atmospheres shown as red lines in Fig. \ref{fig:SEDs}) and the uncertainty of distances obtained from parallax measurements, which were heavily affected by the binary orbital motion \citep{kamath2022GalacticSingles}.

\subsection{Spectroscopic data}\label{ssec:obsspc}
In this subsection, we provide details of the optical and NIR spectral observations used in our study.

\subsubsection{HERMES spectra}\label{sssec:obsspcopt}
The HERMES spectra were obtained as part of a long-term monitoring project (initiated in June 2009 and still ongoing) that made use of the HERMES spectrograph \citep[High Efficiency and Resolution Mercator Echelle Spectrograph;][]{hermes2011} mounted on the 1.2 m Mercator telescope at the Roque de los Muchachos Observatory, La Palma. The project resulted in a substantial collection of high-resolution optical spectra of post-AGB systems, as documented by \cite{winckel2018}.

High-resolution optical spectra from HERMES (R = $\lambda/\Delta\lambda\sim$ 85 000) covered a wavelength range from about 377 nm to 900 nm. The data were reduced through a dedicated pipeline with standard settings as described in \cite{hermes2011}. To observe the whole pulsation cycle of SZ~Mon, a total of 88 spectra were obtained (see Table~\ref{tabA:szmvis}), with a total time span of 3700 days. Similarly, a total of 83 spectra over 4000 days were obtained for DF~Cyg (see Table~\ref{tabA:dfcvis}). 

\subsubsection{APOGEE spectra}\label{sssec:obsspcnir} % 66_5_57_5_48 nm for spectral regions
In this study, we also used NIR spectra from the APOGEE survey, which was a large-scale stellar spectroscopic survey conducted in the NIR regime \cite[H-band;][]{2017AJ....154...94M}. APOGEE made use of two 300-fiber cryogenic spectrographs mounted in different hemispheres. The Northern Hemisphere was studied by the 2.5-metre Sloan Foundation Telescope and the 1-metre NMSU Telescope (for several bright sources) at Apache Point Observatory (APO) in New Mexico, United States. The Southern Hemisphere was observed by the 2.5-metre Ir\'{e}n\'{e}e du Pont Telescope of Las Campanas Observatory (LCO) in Atacama, Chile. APOGEE obtained the medium-resolution (R = 22\,500) spectra across the entire Milky Way as part of the Sloan Digital Sky Survey (SDSS).

SDSS-IV DR17 is the final release of APOGEE data. This release provided spectra, radial velocities, and stellar atmospheric parameters along with individual elemental abundances for more than $657\,000$ stars \citep{2015AJ....150..173N, 2022ApJS..259...35A}. In DR17, three types of APOGEE output files could be accessed: \texttt{apVisit/asVisit} (individual visit raw spectra), \texttt{apStar/asStar} (radial velocity-corrected individual visit and combined spectra), and \texttt{aspcapStar} (fully prepared spectra analysed by ASPCAP, see Section~\ref{sec:int}).

As mentioned by \cite{APOGEE-AGB2020}, ASPCAP was not suited to analyse the spectra of dusty giant stars, so the ASPCAP results (i.e., atmospheric parameters and elemental abundances) for post-AGB and post-RGB stars were out of corresponding grid bounds. Moreover, continuum normalisation routine used by ASPCAP performed poorly for our targets. As a result, we made the decision to omit the \texttt{aspcapStar} files from our study.

Instead, we performed our own spectral analyses (see Section~\ref{sec:san}) where we adopted the \texttt{apStar/asStar} files, which were found to be the most suitable choice for our study. The \texttt{apStar/asStar} files provided APOGEE spectra with a logarithmically-spaced wavelength scale: a common spacing of $\log\lambda_{i+1}-\log\lambda_{i} = 6\times10^{-6}$, starting from 1510.0802 nm.

We note that APOGEE wavelength scale was calibrated using vacuum wavelengths, while we used the NIR line list with transition data provided in standard temperature and pressure (S.T.P.). This lead to a wavelength difference between the spectra and NIR line lists on the order of $\approx0.45$ nm. As stated in APOGEE manual\footnote{Accessible via \url{https://www.sdss4.org/dr17/irspec/}}, the vacuum-to-air wavelength conversion was done according to \cite{1996ApOpt..35.1566C} given by
\begin{multline}
    \lambda_{\rm AIR} = \dfrac{\lambda_{\rm VAC}}{1 + \dfrac{5.792105\times10^{-2}}{238.0185 - [\dfrac{1000}{\lambda_{\rm VAC}}]^2} + \dfrac{1.67917\times10^{-3}}{57.362 - [\dfrac{1000}{\lambda_{\rm VAC}}]^2}},
\end{multline}
where $\lambda_{\rm AIR}$ and $\lambda_{\rm VAC}$ represent air and vacuum wavelengths in nm, respectively.

In addition to arrays of fluxes and flux errors, the \texttt{apStar/asStar} files also included calibration arrays (i.e., sky emission, telluric absorption lines, cosmic rays), which we used to remove the bad pixels from spectra before the analysis (see Section~\ref{ssec:saneis}).

\subsubsection{Epoch selection}\label{sssec:epochs}
Since our study focused on Type II Cepheid variables (SZ~Mon is a W Vir variable and DF~Cyg is an RV Tau variable, see Section~\ref{sec:tar}), we encountered an extra challenge in analysing their spectra because the atmospheric parameters of pulsating variables (i.e., $T_{\rm eff}$, $\log g$, and $\xi_{\rm t}$) experience notable variations throughout the pulsation cycle \citep{manick2019RVTauDFCyg}.

\begin{figure}
    \centering
    \includegraphics[width=.99\linewidth]{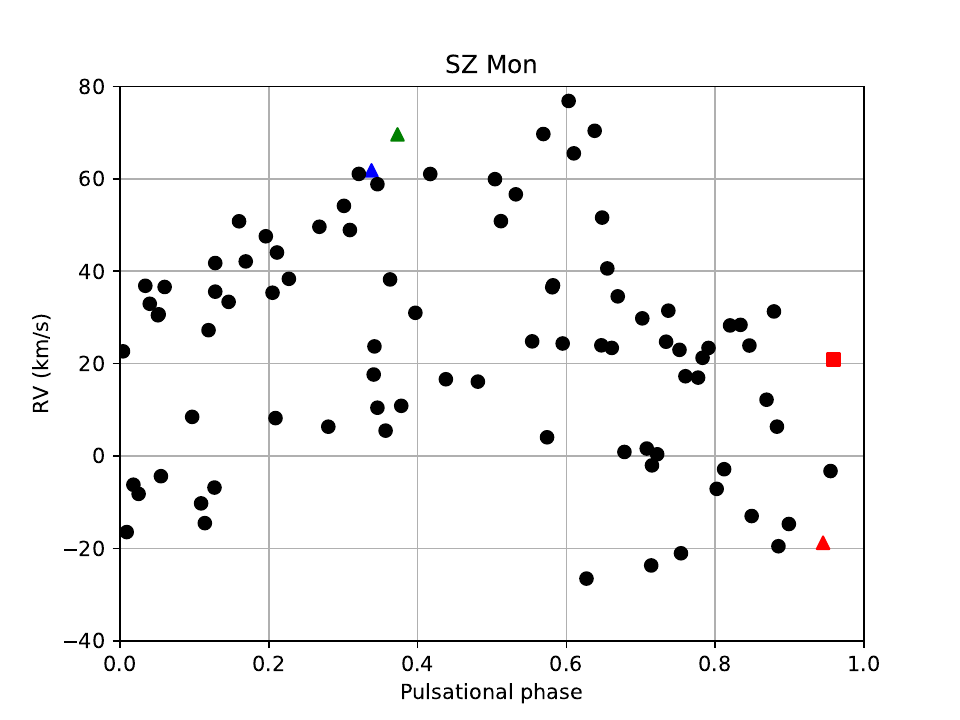}
    \includegraphics[width=.99\linewidth]{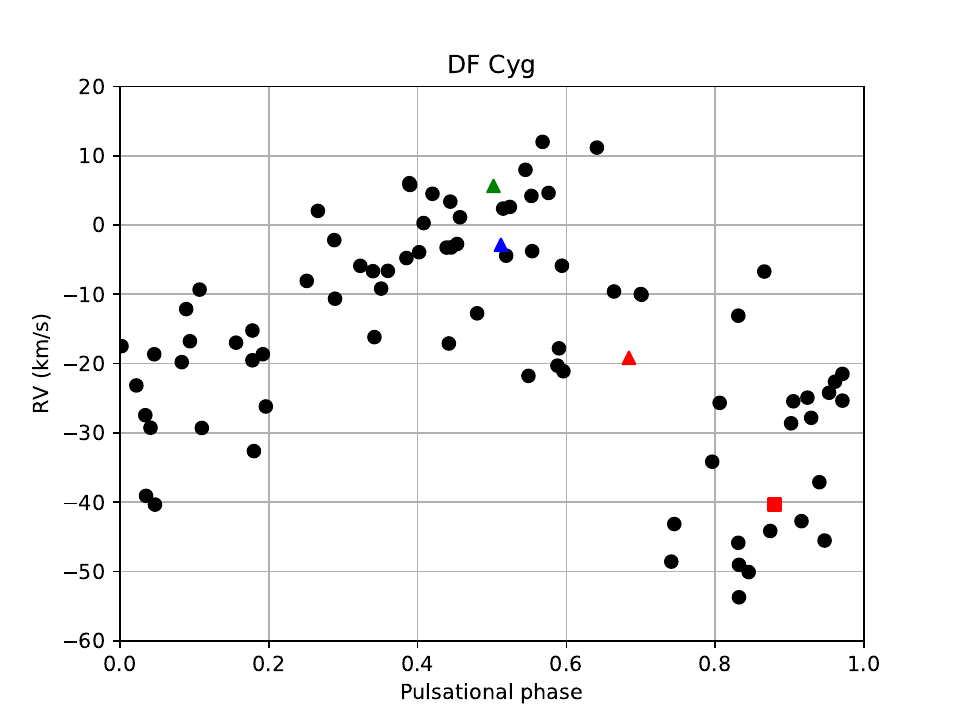}
    \caption{Radial velocity vs pulsation phase plots for SZ Mon (top; $P_\text{puls}$=16.336d) and DF Cyg (bottom; $P_\text{puls}$=25.04d). Green and blue triangles mark the spectra taken at close cooler phases of pulsations: near-infrared APOGEE visit and the optical HERMES visit, respectively. Red markers show the optical HERMES visits obtained at hotter phases of pulsations: triangle for the visit we use in our chemical analysis and square for the visit we use to test the obtained results. For more details see Section~\ref{sssec:epochs}).}\label{fig:RVs}
\end{figure}

Therefore, for each target we needed to ensure that the analysed optical and NIR spectra were observed at close pulsation phases. To do this, we acquired the radial velocities for each spectral visit of each target (see Tables~\ref{tabA:szmvis} and \ref{tabA:dfcvis}): for optical spectra, the radial velocities were derived using E-iSpec (see Section~\ref{ssec:saneis}), while for NIR spectra, we adopted the radial velocities derived by APOGEE's pipeline (see Section~\ref{sssec:obsspcnir}). The acquired radial velocities were then plotted against the corresponding pulsation phases, and the visits for the joint analysis were selected (depicted with blue and green triangles in Fig.~\ref{fig:RVs} for optical and NIR visits, respectively). To validate and extend the number of studied chemical species, we additionally selected two more optical visits for each target at hotter pulsation phases (depicted with red triangle and square in Fig.~\ref{fig:RVs} for main and secondary optical visits, respectively). The comprehensive explanation of epoch selection is provided in Appendix~\ref{app:epo}.

To summarise, based on the pulsation phases and S/N of the visits, we selected the following visits (see Table~\ref{tab:obslog}):
\begin{enumerate}
    \item the optical HERMES spectra to determine atmospheric parameters and elemental abundances from atomic lines: SH\#73, SH\#29, and SH\#47 for SZ~Mon; DH\#83, DH\#26, and DH\#54 for DF~Cyg\footnote{We note that we adopted the following naming convention: the first letter stands for the target (``S'' for SZ~Mon and ``D'' for DF~Cyg), the second letter stands for the instrument (``H'' for HERMES and ``A'' for APOGEE), and the number is the chronological order of the visit.};
    \item the NIR APOGEE spectra to determine elemental abundances and carbon isotopic ratios from molecular bands: SA\#2 for SZ~Mon; DA\#1 for DF~Cyg.
\end{enumerate}

\begin{table*}
    \centering
    \caption{The details of spectroscopic observations of SZ Mon and DF Cyg. The choice of shown visits and visit naming convention are explained in Section \ref{sssec:epochs}. For the full observing log, see Tables~\ref{tabA:szmvis} and \ref{tabA:dfcvis}).}\label{tab:obslog}
    \begin{tabular}{ccccccc}\hline
         Visit & Obs. date & Obs. start & Obs. ID & Instrument + & Radial velocity & Pulsational\\
         number & ~ & (UT) & ~ & telescope & (km/s) & phase \\ \hline
         \multicolumn{7}{c}{SZ Mon (2MASS J06512784-0122158)} \\ \hline
         SH\#73$^{1}$ & 2018-02-18 & 01:43:40.80 & 00866282 & HERMES + Mercator$^{a}$ & --18.78$\pm$0.05 & 0.95 \\
         SH\#29$^{2}$ & 2012-03-24 & 22:40:48.00 & 00397261 & HERMES + Mercator$^{a}$ & 20.84$\pm$0.04 & 0.96 \\
         SH\#47$^{3}$ & 2013-04-10 & 21:04:19.20 & 00458609 & HERMES + Mercator$^{a}$ & 61.82$\pm$0.09 & 0.34 \\
         SA\#2$^{4}$ & 2016-01-19 & 06:24:28.80 & 8803.57406.58 & APOGEE-N + SFT$^{b}$ & 69.60$\pm$0.20 & 0.37 \\ \hline
         \multicolumn{7}{c}{DF Cyg (2MASS J19485394+4302145)} \\ \hline
         DH\#83$^{1}$ & 2020-08-16 & 00:38:52.80 & 00972481 & HERMES + Mercator$^{a}$ & --19.17$\pm$0.04 & 0.68 \\
         DH\#26$^{2}$ & 2012-06-24 & 04:22:04.80 & 00412205 & HERMES + Mercator$^{a}$ & --40.4$\pm$0.2 & 0.88 \\
         DH\#54$^{3}$ & 2014-07-06 & 03:40:19.20 & 00574546 & HERMES + Mercator$^{a}$ & --2.86$\pm$0.06 & 0.51 \\
         DA\#1$^{4}$ & 2016-09-14 & 04:20:38.40 & 9129.57645.89 & APOGEE-N + SFT$^{b}$ & 5.64$\pm$0.05 & 0.50 \\ \hline
    \end{tabular}\\
    \textbf{Notes:} $^a$Mercator is the 1.2-metre telescope at the Observatorio del Roque de Los Muchachos, La Palma. $^b$SFT is the 2.5-metre Sloan Foundation Telescope located at Apache Point Observatory, New Mexico. $^{1,2}$Observations were taken at hotter pulsation phases which are ideal for studying atomic abundances (1) and confirming derived atmospheric parameters (2). $^{3,4}$Observations were taken at cooler pulsation phases which are ideal for deriving atmospheric parameters from HERMES spectra (3) and passing them to the analysis of molecular abundances from APOGEE spectra (4).
\end{table*}

\section{Spectral analysis}\label{sec:san}
To determine the atmospheric parameters, elemental abundances, and the carbon isotopic ratios for evolved stars, we developed E-iSpec -- a modified version of iSpec \citep{2014A&A...569A.111B, 2019MNRAS.486.2075B}. Specifically, our modifications focused on improving the accuracy and reliability of elemental abundance measurements for objects with complex atmospheric conditions, and on adding the isotopic ratios derivation to the spectral analysis.

As a validation of our approach for epoch selection, we performed the chemical analysis for all selected visits of SZ~Mon and DF~Cyg (three optical and one NIR visits for each target, see Section~\ref{sssec:epochs}). We found that the differences in atmospheric parameters and elemental abundances were generally smaller than the corresponding uncertainties (except an expected change in the effective temperatures; see Table~\ref{tabA:tststp}).

In the following subsections, we briefly present iSpec and its limitations. Subsequently, we introduce E-iSpec and present the analyses and results for our target stars: SZ~Mon and DF~Cyg.

\subsection{iSpec and its limitations}\label{ssec:sanisp}
iSpec is an integrated spectroscopic software framework, which is capable of determining atmospheric parameters such as effective temperature, surface gravity, metallicity, micro/macroturbulence, rotation and individual elemental abundances for AFGKM stars. The full details of iSpec are presented in \cite{2014A&A...569A.111B, 2019MNRAS.486.2075B}. In brief, iSpec integrates MARCS \citep{2008A&A...486..951G} and different ATLAS9 \citep[i.e., Kurucz, Castelli \& Kurucz, KuruczODFNEW;][]{2003IAUS..210P.A20C} one-dimensional (1D) model atmospheres with assumed local thermodynamic equilibrium (LTE) together with the following radiative transfer codes: SPECTRUM \citep{1994AJ....107..742G}, Turbospectrum \citep{1998A&A...330.1109A, turbospec2012ascl}, SME \citep{1996A&AS..118..595V}, MOOG \citep{1973PhDT.......180S}, and Synthe/WIDTH9 \citep{1993KurCD..18.....K, 2004MSAIS...5...93S}. These radiative transfer codes are implemented in iSpec to work in two different approaches: equivalent widths (EW) method and synthetic spectral fitting (SSF) technique. Furthermore, to carry out the spectral analyses, iSpec provides several line lists with a wide wavelength coverage (overall, from 300\,nm to 4\,$\mu$m). The line lists included are: the VALD3 line list \citep{kupka2011vald}, the APOGEE line list \citep{2021AJ....161..254S}, the Gaia-ESO line list \citep{heiter2021GaiaESO}, the ATLAS9 line lists \citep{kurucz1995linelist}, and the original SPECTRUM line list, which contains atomic and molecular lines obtained mainly from the NIST Atomic Spectra Database \citep{ralchenko2005nist}. Additionally, iSpec provides a collection of ready-to-use solar abundances from \cite{anders1989solar, grevesse1998solar, asplund2005solar, grevesse2007solar, asplund2009}.

We note that iSpec offers two usability modes: a GUI version and a Python version. E-iSpec (see Section~\ref{ssec:saneis}) exclusively utilises the Python version, which offers enhanced functionality.

While iSpec is capable of the spectral analysis of a wide range of targets, we note that it is not specifically designed for our goal. Firstly, iSpec routine for automatic continuum normalisation performs poorly for evolved stars, the spectra of which contain prominent molecular bands. Secondly, the molecular line lists presented in iSpec (especially for NIR regime) were not up-to-date. Thirdly, the error estimation of elemental abundances in iSpec only takes into account line-to-line (L2L) scatter completely ignoring the systematic errors caused by uncertainties of atmospheric parameters. Finally and most importantly, the isotopic ratios were not addressed in chemical analysis routines of iSpec. To overcome these limitations, we developed a Python wrapper for iSpec: E-iSpec.

\subsection{E-iSpec: Evolved stars solution of iSpec}\label{ssec:saneis}
E-iSpec\footnote{Accessible via \url{https://github.com/MaksTheUAstronomer/E-iSpec.git}} is a semi-automated spectral (optical and NIR) analysis tool implemented in Python, primarily derived from iSpec but enhanced with extended capabilities described below. E-iSpec allows for the determination of atmospheric parameters, elemental abundances (from both atomic lines and molecular lines), and isotopic ratios, particularly for evolved stars with complex atmospheres.

E-iSpec uses 1D LTE model atmospheres (from MARCS and ATLAS9 grids)\footnote{We note that spherically-symmetric MARCS model atmospheres generally perform better for cool giants \citep{models2012apogee}.} and relies on the following radiative transfer codes, which were chosen due to their performance and reliability: Turbospectrum \citep{turbospec2012ascl} for synthetic spectral fitting technique and Moog \citep{1973PhDT.......180S} for equivalent width method. Line lists for atomic and molecular transitions, as detailed in Appendix~\ref{app:add}, are drawn from VALD3 for optical spectra ($\lambda\sim300-1100$ nm) and DR17 APOGEE for NIR spectra ($\lambda\sim1500-1700$ nm) \citep{kupka2011vald, hayes2022bacchusapogeelinelist}. Solar abundances are derived from LTE values provided by \cite{asplund2009}.

In summary, E-iSpec allows to prepare the stellar spectra for chemical analysis and to determine atmospheric parameters along with elemental abundances and isotopic ratios. In the following subsections, we describe the capabilities of E-iSpec. The final atmospheric parameters, elemental abundances, and carbon isotopic ratio for SZ~Mon and DF~Cyg derived with E-iSpec are provided in Tables~\ref{tab:fnlabsSZM} and \ref{tab:fnlabsDFC}, respectively. Additionally, we present an extensive benchmarking of E-iSpec in Appendix~\ref{app:tst}.

\subsubsection{Preparing data for spectral analyses}\label{sssec:saneisprep}
To prepare the spectra for the atmospheric parameter and elemental abundance analyses, E-iSpec offers a series of rigorous data pre-processing steps. These includes continuum normalisation, cosmic rays removal, radial velocity correction, and construction of the initial line list that is used in the stellar spectral analyses.

%(the latter was conducted with consideration of the expected atmospheric parameters specific to the observation in question, as described in the previous subsection).

In E-iSpec, we provided two methods of continuum normalisation: manual and automatic. The manual normalisation involved fitting $10^{\rm th}$-order polynomial functions through interactively determined continuum points, typically spaced at intervals of $\approx$10\,nm. In contrast, the automatic normalisation involved fitting $3^{\rm rd}$-order splines through automatically determined continuum points. The continuum points in the automatic normalisation routine were determined using the iSpec functions \texttt{fit\_continuum} and \texttt{normalize\_spectrum}. In brief, \texttt{fit\_continuum} function performed a median filtering for every 3 data points and a maximum filtering for every 30 data points to the spectral data. The remaining data points were then fitted with the number of spline knots automatically determined by the normalisation routine. The selection of a normalisation procedure depended on the characteristics of the spectral features. In general, spectra containing atomic lines tended to yield the best results when a two-step approach is applied, involving manual normalisation followed by automatic normalisation. On the other hand, for spectra featuring molecular bands, manual continuum normalisation alone was recommended.  

Once the spectra were normalised, the cosmic rays could be removed using the iSpec function \texttt{create\_filter\_cosmic\_rays} with default settings (\texttt{resampling\_wave\_step} was the wavelength step, \texttt{window\_size} was 15, and \texttt{variation\_limit} was 0.5).

Then, the normalised and cleaned spectra were corrected for the radial velocity using the iSpec function \texttt{cross\_correlate\_with\_template}. In case of the HERMES spectra, the correction started with the spectrum smoothing by degrading its resolution from 85~000 to 50~000, and then cross-correlating the smoothed spectrum with the Arcturus template spectrum to determine the radial velocity shift. This step ensured precise radial velocity determination even for low-S/N ($<70$) spectra. For the APOGEE spectra, the radial velocity correction has already been performed by the APOGEE pipeline (see Section~\ref{ssec:obsspc}). The absolute difference of radial velocities between our method and the APOGEE pipeline was found to be below 1 km/s.

Next, we employed the automatic line identification in iSpec (\texttt{fit\_lines} with \texttt{max\_atomic\_wave\_diff} parameter being 0.005). This identification heavily depended on expected line strengths (defined as `theoretical depth' and `theoretical EW'), which were calculated for the initially specified set of the atmospheric parameters. Consequently, the closer the input atmospheric parameters were to the final results, the better spectral lines were identified in spectra. For our target sample, this effect was prominent only between visits obtained at slightly hotter and cooler phases (see Table~\ref{tab:obslog}).

\begin{figure*}
    \centering
    \includegraphics[height=.49\linewidth, angle=-90]{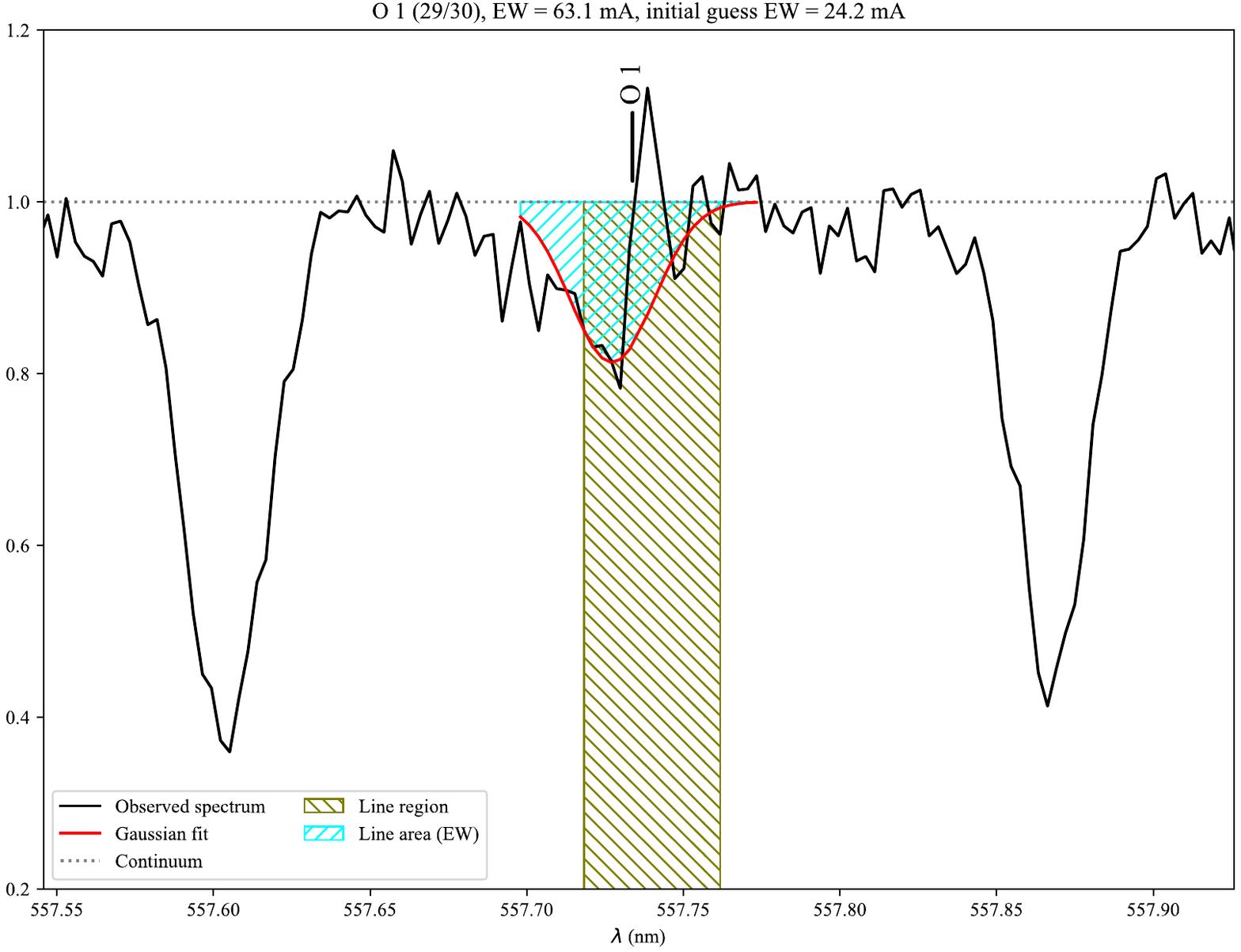}
    \includegraphics[height=.49\linewidth, angle=-90]{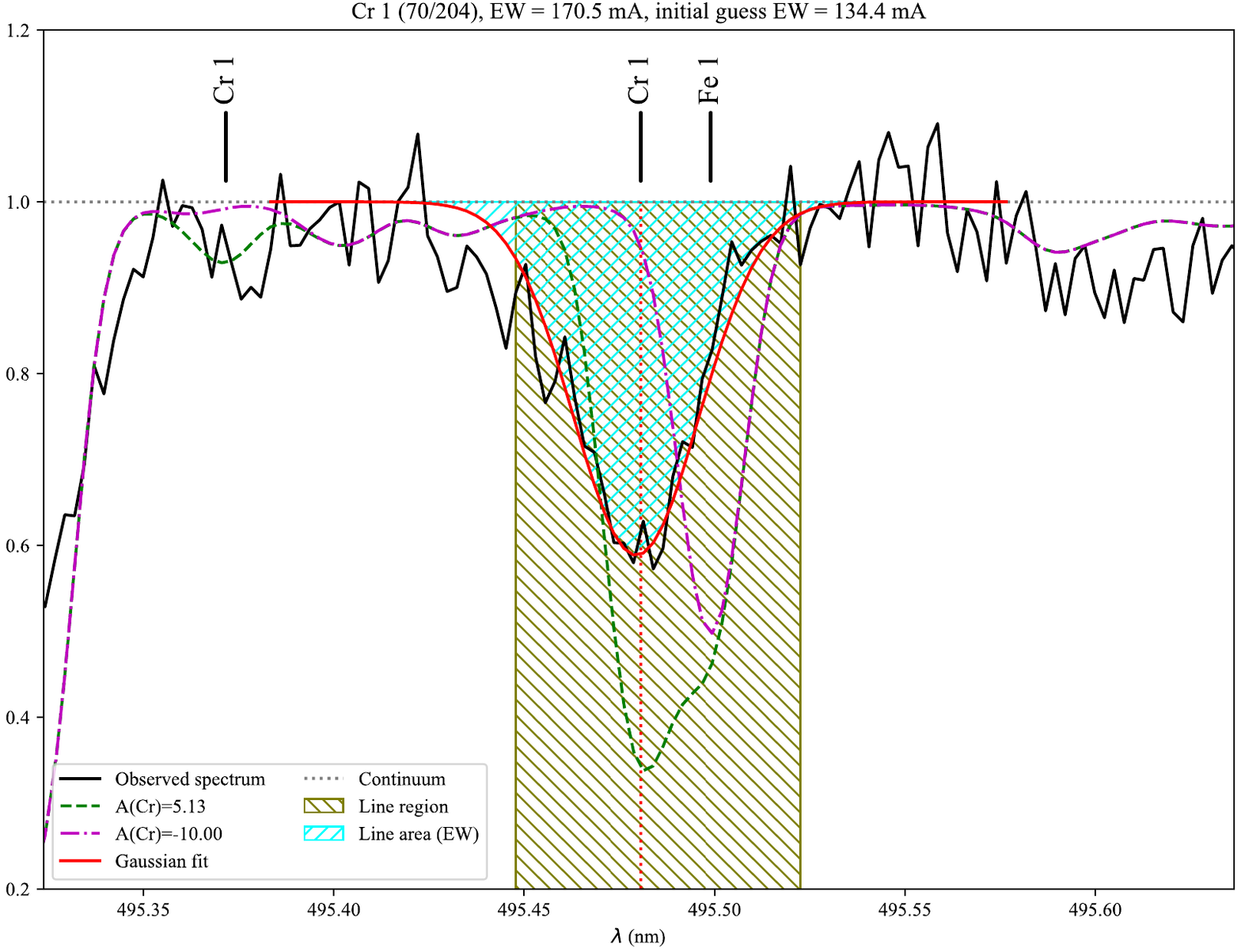}
    \includegraphics[height=.49\linewidth, angle=-90]{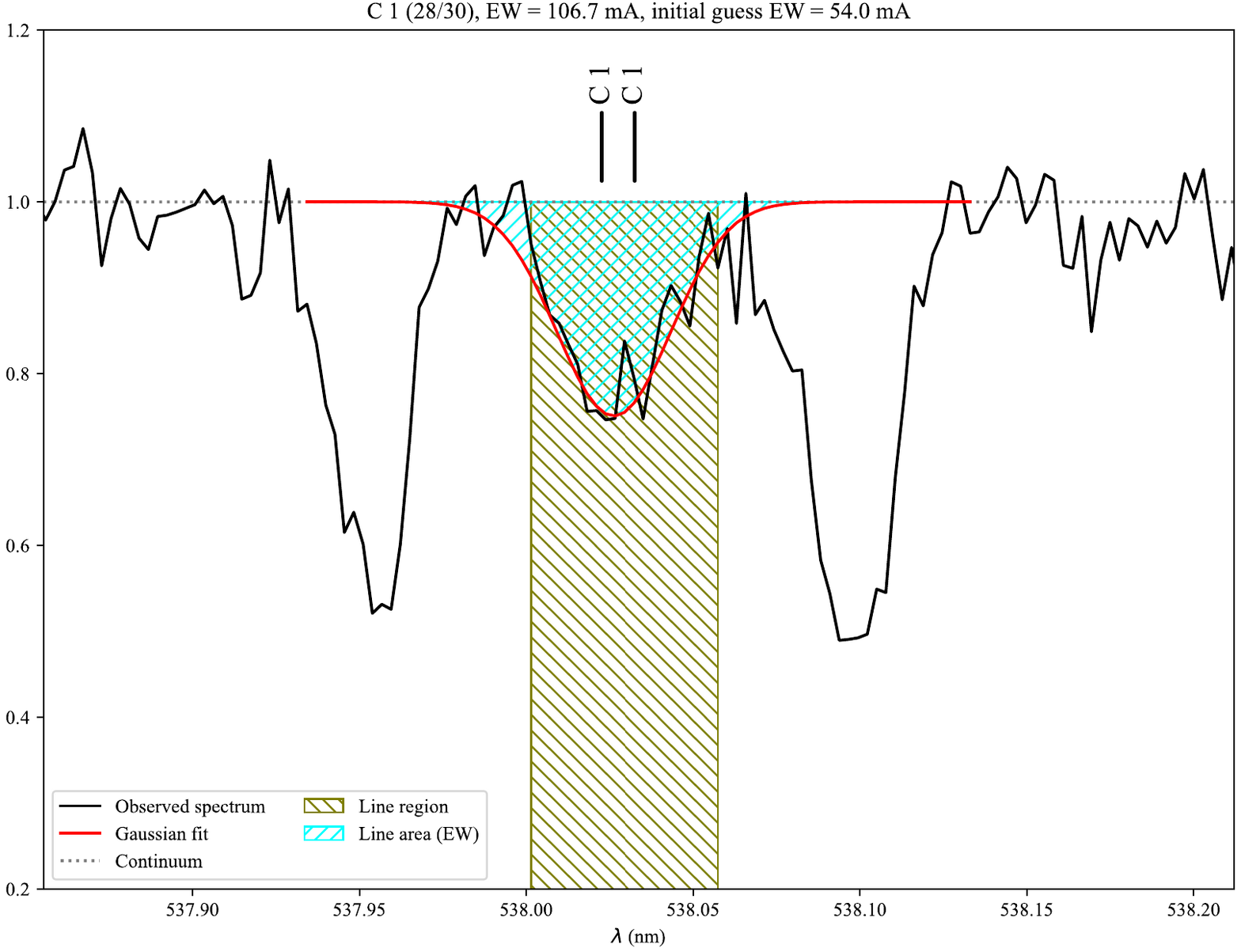}
    \includegraphics[height=.49\linewidth, angle=-90]{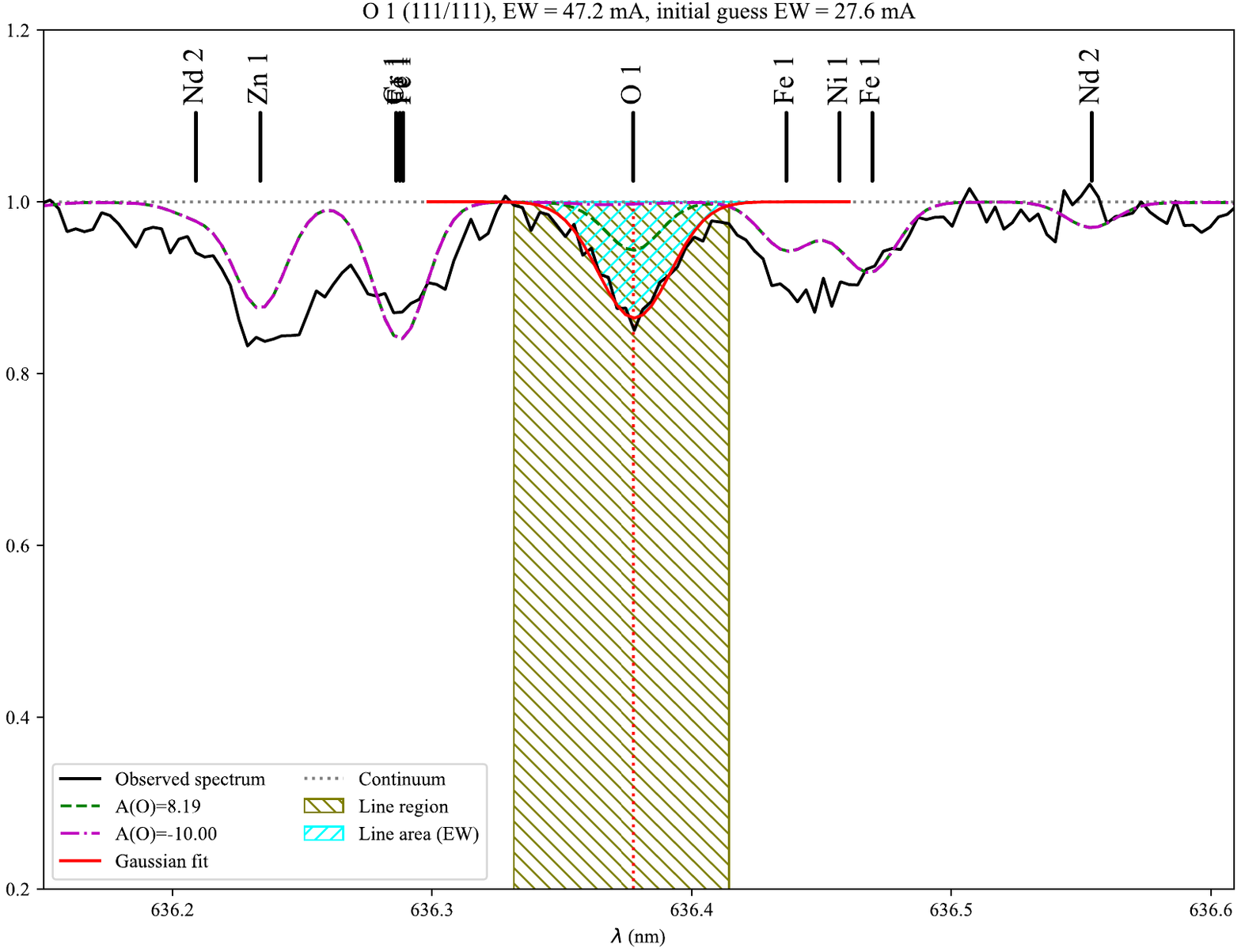}
    \caption{The examples of the spectral lines in different spectra of SZ~Mon showing the process of manual line selection: from the automatically identified spectral lines we removed spectral lines affected by the cosmic hit (top left panel), blended spectral lines (top right panel), and ``self-blended'' spectral lines (bottom left panel). A typical spectral line used for the chemical analysis is shown in the bottom right panel. The legend for the symbols and colours used are included within the plot.}\label{fig:linsel}
\end{figure*}

Finally, we manually selected lines from the automatic line list (see Fig.~\ref{fig:linsel}). The process of line selection involved three steps: i) coarse filtering of poorly identified spectral lines (top left panel), ii) fine filtering of blended spectral lines (top right panel), and iii) final filtering of outliers (bottom left panel). To perform the fine filtering, we overplotted the observed spectrum with two synthetic spectra, which have atmospheric parameters fixed at literature values and most abundances fixed at metallicity-scaled solar values. The difference between synthetic spectra lied in the abundance of the element identified for the selected spectral feature: for one spectrum, $[$X/H$]$ was set at -10 dex to simulate the absence of the element; for the other spectrum, $[$X/H$]$ was set at metallicity-scaled solar value. We note that, while applying the first two filtering steps aided in excluding most of the misidentified lines and blends, the third step was needed to remove the blends of the element with itself (both same and different ionisation). Once the automatically selected lines were manually filtered (bottom right panel), the final line list was used to derive atmospheric parameters and elemental abundances.

\subsubsection{Determining atmospheric parameters and elemental abundances from atomic lines}\label{sssec:stepar}
To determine the atmospheric parameters ($T_{\rm eff}$, $\log g$, $[$Fe/H$]$, and $\xi_{\rm t}$) we used iSpec function \texttt{model\_spectrum\_from\_ew} with the following settings: radiative transfer code was chosen to be Moog, iterations were run until convergence with a maximum number of 10 (as proposed by \cite{2014A&A...569A.111B, 2019MNRAS.486.2075B}), and automatic detection of outliers was almost switched off by setting the sigma clipping level at 10-$\sigma$. This function used EW method by comparing the observed spectral lines with the Gaussian fits. The errors of atmospheric parameters provided by this function were calculated from the covariance matrix constructed by the non-linear least-squares fitting algorithm.

The elemental abundances ($[$X/H$]$) were determined from atomic lines using function \texttt{model\_spectrum} with similar settings: Moog was the preferred radiative transfer code, and the iterations were run until convergence with a maximum number of 10. This function provided the errors of elemental abundances as standard deviations of the measured values (line-to-line scatter), so we needed to also account for the errors caused by the atmospheric parameters (systematic error; see Section~\ref{sssec:sanerr}). In Table~\ref{tabA:tstsmp} we provided the comparison of atmospheric parameters of chemically diverse sample of post-AGB stars: literature values were obtained from the manually selected spectral lines, while our values are obtained using automatic EW method.

\subsubsection{Deriving elemental abundances and isotopic ratios from molecular bands}\label{sssec:abuiso}
To derive the carbon isotopic ratios, we developed our own algorithm, which we discuss below. To test its performance, we expanded it to also derive elemental (atomic and molecular) abundances, which we could detect in APOGEE spectra (see Section~\ref{sssec:obsspcnir}). We note that our SSF algorithm for molecular bands was generally slower than iSpec's SSF routine for atomic lines. However, we benchmarked our algorithm for red giants (\cite{masseron2019APOGEE+BACCHUS}; see Table~\ref{tabA:tstabs}) and for different visits of SZ~Mon and DF~Cyg (optical counterparts were obtained using EW method; see Table~\ref{tabA:tststp}).

We started the abundance analysis with identification of spectral regions that were sensitive to specific elements and isotopes. For this, we synthesised a reference spectrum with all elemental abundances set to metallicity-scaled solar (MSS) values. We then created individual synthetic spectra for each element, with that element's abundance enhanced by 1 dex above MSS value while keeping all other elements at MSS values. Subtracting the reference spectrum from each enhanced spectrum resulted in a plot of spectral sensitivity for each selected chemical element. We note that for the degenerate broadening effects (spectral resolution, rotational velocity, and macroturbulent velocity), we used fixed values for resolution (R = 22500) and rotational velocity ($v\sin i$ = 0 km/s), and only the macroturbulence $v_{\rm mac}$ was a free parameter, accounting for all broadening effects.

We then manually filtered the regions in these spectral sensitivity plots where flux differences exceeded the overall standard deviation of the normalised flux in the corresponding enhanced spectrum. Upon this filtering, we obtained the final spectral regions used for abundance analysis.

Once we identified the final spectral windows, we synthesised spectra for different element abundances and used a $\chi^2$ minimisation procedure to determine the best-fit elemental abundances and isotopic ratios. In the current version of the script, we found the minimal $\chi^2$ values using the Fibonacci search method, which has a time complexity of $O(\log N)$. The $\chi^2$ for the selected spectral window is given by
\begin{equation}
    \chi^2 = \sum_\lambda\frac{(O_\lambda-S_\lambda)^2}{\sigma_S^2},
\end{equation}
where $O_\lambda$ and $S_\lambda$ represent the normalised fluxes of the observed and synthetic spectra, respectively. $\sigma_S$ represents the standard deviation of the synthetic flux within a specific spectral feature. This calculation is conducted across the wavelength range outlined in the NIR line lists, which are detailed in Tables~\ref{tabA:nirlstszm} and \ref{tabA:nirlstdfc} within Appendix~\ref{app:add}. These tables provide multicolumn wavelength ranges corresponding to different spectral features we used. The squared fractions ($\frac{(O_\lambda-S_\lambda)^2}{\sigma_S^2}$) are then summed over the entire wavelength range of the studied feature.

We note that in iSpec, the reference isotopic ratios are input in the SPECTRUM format $p(\textrm{X}_i) = N(\textrm{X}_i)/N(\textrm{X})$, where $N(\textrm{X}_i)$ and $N(\textrm{X})$ are number densities of element X (in specific isotopic form $\textrm{X}_i$ and in total, respectively). Similarly, the common convention for CNO isotopic ratios can be described as
\begin{equation}
    \frac{\textrm{X}_{\rm main}}{\textrm{X}_{\rm var}} = \frac{N(\textrm{X}_{\rm main})}{N(\textrm{X}_{\rm var})} = \frac{N(\textrm{X}_{\rm main})/N(\textrm{X})}{N(\textrm{X}_{\rm var})/N(\textrm{X})} = \frac{p(\textrm{X}_{\rm main})}{p(\textrm{X}_{\rm var})},
\end{equation}
where $\textrm{X}_{\rm main}$ and $\textrm{X}_{\rm var}$ are main and variant isotopic forms of element X, respectively.

We also note that E-iSpec inherits the Turbospectrum method of operating with the total [C/H] abundance for a selected $^{12}$C/$^{13}$C ratio. These two parameters should simultaneously fit the molecular bands of CO and CN, as well as their isotopologues $^{13}$CO and $^{13}$CN, respectively. For atomic lines, $^{12}$C and $^{13}$C contribute to the same spectral features, hence the [C/H] abundances derived from the optical spectra were crucial to validate the [C/H] abundances derived from the molecular spectra.

\begin{figure*}
    \centering
    \includegraphics[width=.49\linewidth]{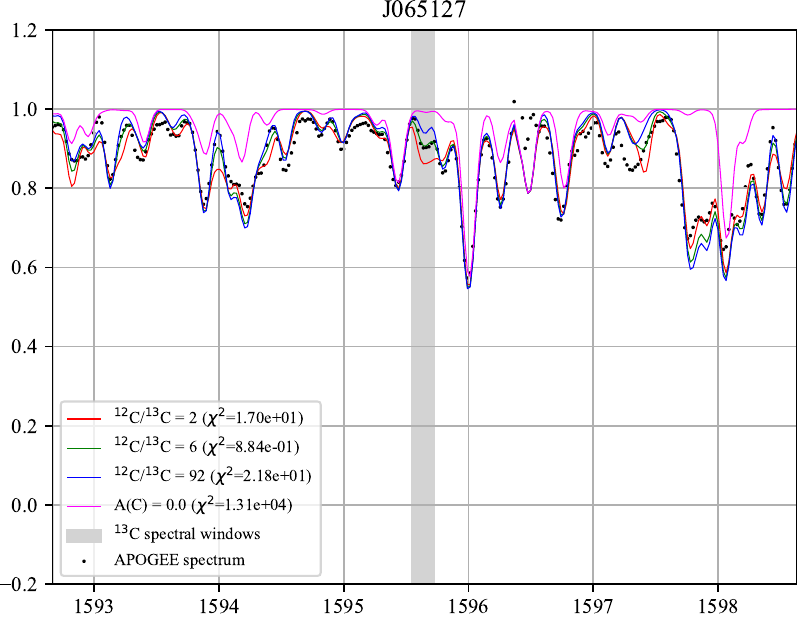}
    \includegraphics[width=.49\linewidth]{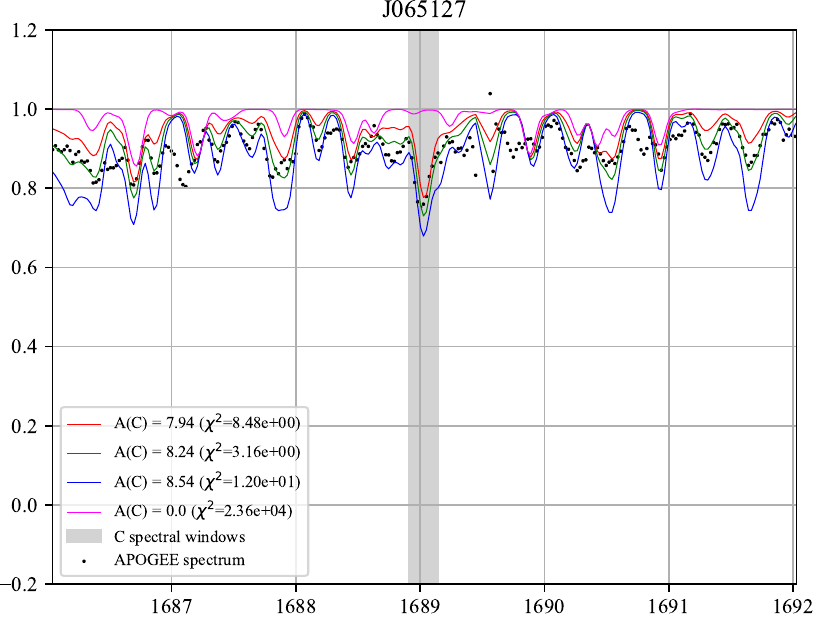}
    \includegraphics[width=.49\linewidth]{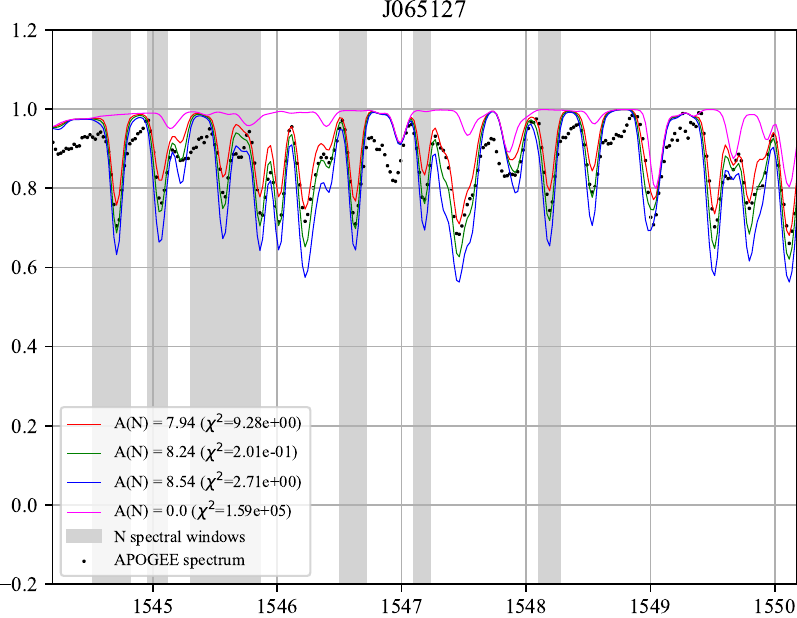}
    \includegraphics[width=.49\linewidth]{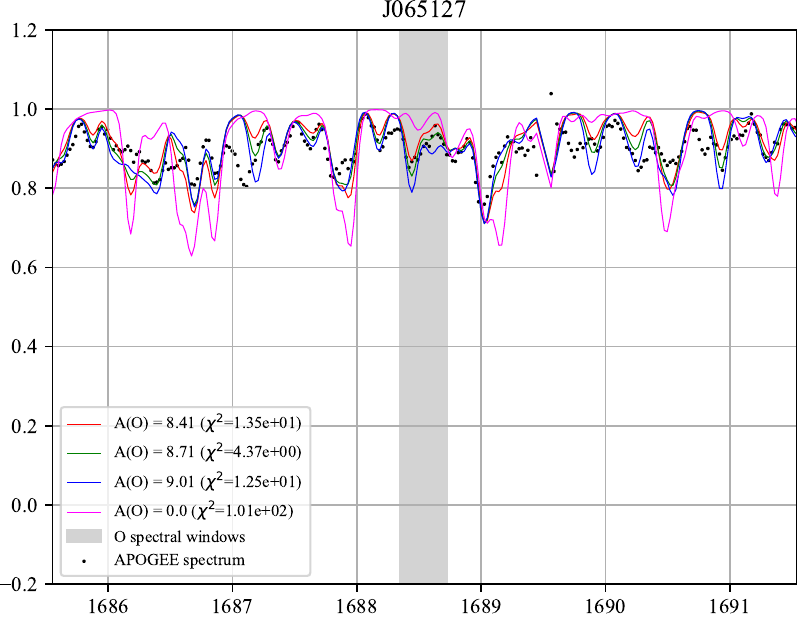}
    \caption{Example fits of spectral features sensitive to $^{12}$C/$^{13}$C (top left panel), [C/H] (top right panel), [N/H] (bottom left panel), and [O/H] (bottom right panel) in the spectrum of SZ~Mon (shown as shaded areas). Observed spectrum is marked with black dots. Synthetic spectra with best-fit--0.3 dex, best-fit, best-fit+0.3 dex, and ``zero'' elemental abundances are graphed with magenta, green, red, and blue lines, respectively (``zero'' abundance is achieved by excluding the spectral features related to the element in question from the line list). Abundances of all other chemical elements in the synthetic spectra are fixed at metallicity scaled solar values. We highlight that the lower right panel demonstrates the well-known inverse relationship between oxygen abundance and the strength of spectral features associated with carbon-bearing molecules such as CN and C$_2$. The legend for the symbols and colours used are included within the plot. For more details see Section~\ref{sssec:abuiso}.}\label{fig:SpWnSZM}
\end{figure*}

\begin{figure*}
    \centering
    \includegraphics[width=.49\linewidth]{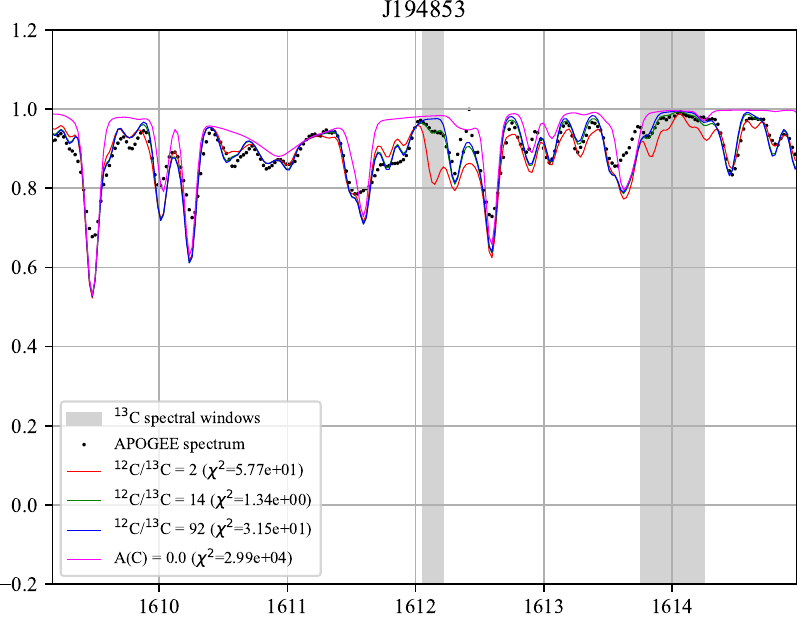}
    \includegraphics[width=.49\linewidth]{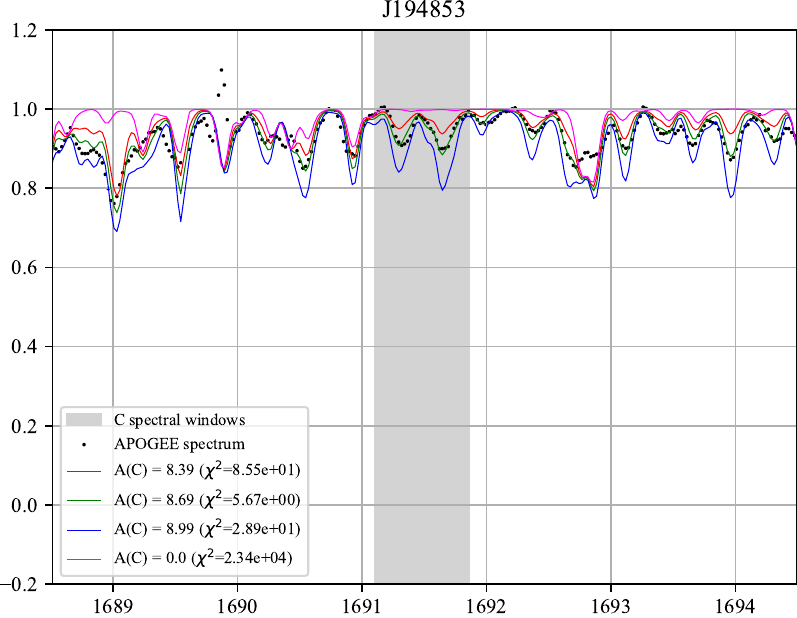}
    \includegraphics[width=.49\linewidth]{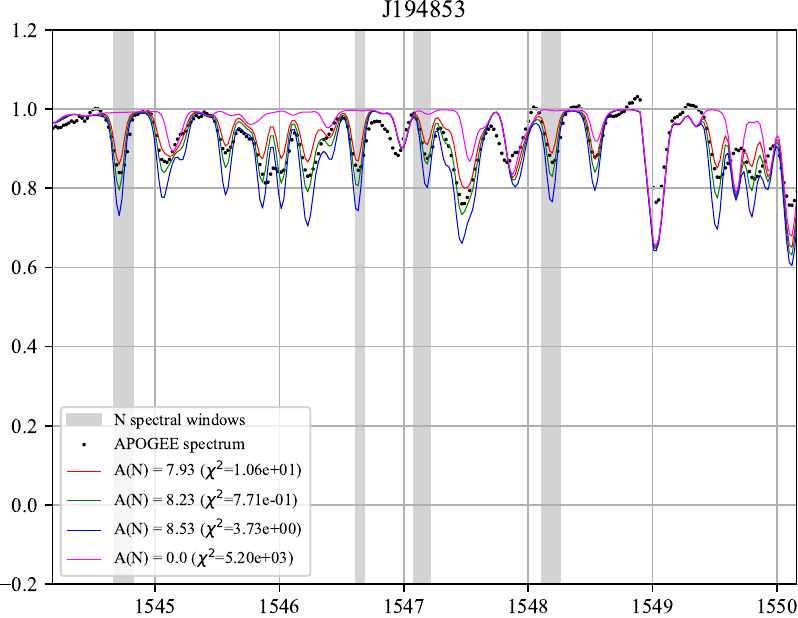}
    \includegraphics[width=.49\linewidth]{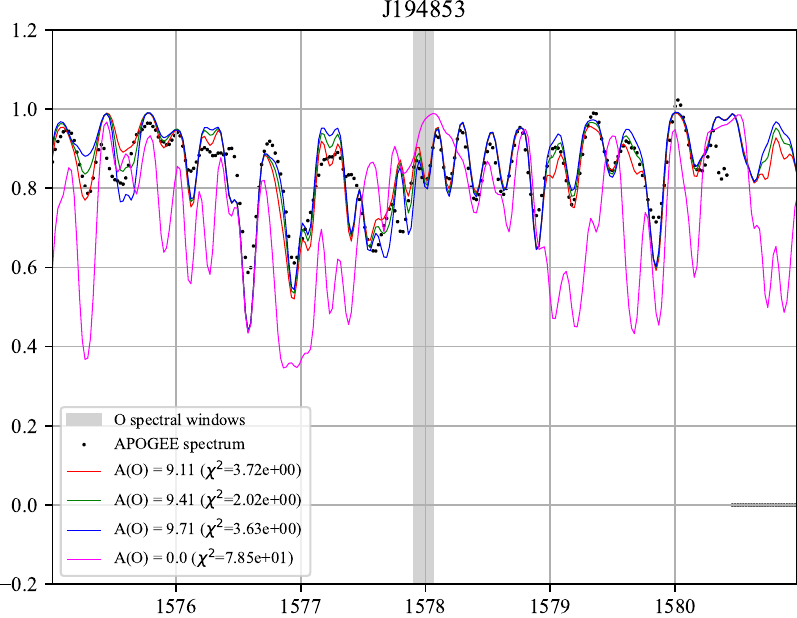}
    \caption{Example fits of spectral features sensitive to $^{12}$C/$^{13}$C (top left panel), [C/H] (top right panel), [N/H] (bottom left panel), and [O/H] (bottom right panel) in the spectra of DF~Cyg (shown as shaded areas). Observed spectrum is marked with black dots. Synthetic spectra with best-fit--0.3 dex, best-fit, best-fit+0.3 dex, and ``zero'' elemental abundances are graphed with magenta, green, red, and blue lines, respectively (``zero'' abundance is achieved by excluding the spectral features related to the element in question from the line list). Abundances of all other chemical elements in the synthetic spectra are fixed at metallicity scaled solar values. We highlight that the lower right panel demonstrates the well-known inverse relationship between oxygen abundance and the strength of spectral features associated with carbon-bearing molecules such as CN and C$_2$. The legend for the symbols and colours used are included within the plot. For more details see Section~\ref{sssec:abuiso}.}\label{fig:SpWnDFC}
\end{figure*}

In Fig.~\ref{fig:SpWnSZM} for SZ~Mon and Fig.~\ref{fig:SpWnDFC} for DF~Cyg, we show examples of spectral regions (windows indicated in grey) sensitive to the $^{12}$C/$^{13}$C ratio (top left panel), [C/H] (top right panel), [N/H] (bottom left panel), and [O/H] (bottom right panel). The observed spectra are shown with black dotted lines, while the synthetic spectra are shown with pink, blue, green, and red solid lines. The synthetic spectra are calculated in the assumption of metallicity scaled solar abundances of all elements except the one in question. For $^{13}$C, pink spectra were synthesised for the assumed absence of the carbon (by removing all carbon-containing features from the line list), blue and red spectra set the boundaries of the solution search for $^{12}$C/$^{13}$C ratio (solar value of 92 and exotic value of 2, respectively), and green spectra were synthesised for the best-fit value of $^{12}$C/$^{13}$C ratio. For each of CNO elements, the spectra were synthesised for different elemental abundances: conditional absence of the element (pink; the abundance is set at the best-fit value, but the molecular bands containing this element and its atomic lines are excluded from the line list), best-fit decreased by 0.3 dex (blue), best-fit (green), and best-fit increased by 0.3 dex (red) abundances, respectively.

In case of DF~Cyg, there was one additional step in our analysis of CNO molecular bands. Given the near-solar metallicity of this target, spectral features in spectra of DF~Cyg (i.e., atomic lines and molecular bands) were significantly blended (compared to the spectra of more metal-poor SZ~Mon). This high level of blending lead to a degeneracy between T$_{\rm eff}$, $[$C/H$]$, and $[$O/H$]$, which was disentangled by running the iterative minimisation process: for different temperatures we fixed $[$C/H$]$ and $[$O/H$]$ in turns until all three parameters converged. We tested the range of temperatures from 4\,400\,K to 5\,400\,K (where we expected the solution to be, based on visual inspection of the aforementioned molecular bands) with increments of 100\,K, which were indicative of our uncertainties in T$_{\rm eff}$. As a result, this iteration process converged at T$_{\rm eff}$ = 4\,500\,K for DA\#1 and at T$_{\rm eff}$ = 5\,000\,K for DA\#2, highlighting our choice of APOGEE visit.

\subsubsection{Error estimation for atmospheric parameters and elemental abundances}\label{sssec:sanerr}

For the atmospheric parameters, we relied on iSpec's built-in error estimation (see Section~\ref{ssec:sanisp}).

For the elemental abundances, the error estimation procedure involved analysing the errors, which arose from the scattering of data points between spectral lines (random errors) and the uncertainties caused by the uncertainties of atmospheric parameters, such as temperature $T_{\rm eff}$, surface gravity $\log g$, metallicity $[$Fe/H$]$, and microturbulent velocity $\xi_{\rm t}$ (systematic errors).

We followed the procedure explained in the appendix of \cite{mcwilliam1995errors} with the assumption that atmospheric parameters are independent \citep{2014A&A...563L...5D}. In short, the overall abundance uncertainty was determined by taking the quadratic sum of errors arising from line-to-line scatter ($\sigma_{\rm l2l}/\sqrt{N_l}$ where $N_l$ was the number of lines of an ionisation level of an element in question) and from atmospheric parameter uncertainties ($\sigma_{T_{\rm eff}}$, $\sigma_{\log g}$, $\sigma_{\xi_{\rm t}}$), given as
\begin{equation}
    \sigma_{\rm tot, [X/H]} = \sqrt{\left(\frac{\sigma_{\rm l2l}}{\sqrt{N_{l}}}\right)^2 + \left(\sigma_{T_{\rm eff}}\right)^2 + \left(\sigma_{\log g}\right)^2 + \left(\sigma_{\xi_{\rm t}}\right)^2}.
\end{equation}

We note that in cases when only one spectral line of an element was used for abundance derivation, we conventionally used 0.2 dex as an upper limit of the uncertainty.

\subsection{Spectral analysis results for SZ Mon and DF Cyg}\label{ssec:sanabs}
In this study, we used E-iSpec to determine atmospheric parameters, elemental abundances, and carbon isotopic ratios for two binary post-RGB stars: SZ~Mon and DF~Cyg. We found that the derived atmospheric parameters and elemental abundances are in good agreement with the previous studies of SZ~Mon \citep{maas2007t2cep} and DF~Cyg \citep{giridhar2005rvtau}, which is expected given the proximity of pulsation phases in our study and the corresponding literature (see Tables~\ref{tab:fnlabsSZM} and \ref{tab:fnlabsDFC}, as well as Fig.~\ref{fig:fnlabs}). We calculated more accurate elemental abundances and increased the amount of analysed species for SZ~Mon (9 newly studied species) and DF~Cyg (8 newly studied species), including $^{13}$C.

Similarly to \cite{maas2007t2cep} and \cite{giridhar2005rvtau}, we detected the chemical depletion profiles for both targets. To quantitatively describe the chemical depletion, we used [Zn/Ti] abundance ratio and the turn-off temperature $T_{\rm turn-off}$ \citep{oomen2019depletion, kluska2022GalacticBinaries}. Due to its volatile nature, Zn \citep[$T_{\rm cond,~Zn}=726$ K; ][]{lodders2003CondensationTemperatures} is anticipated to exhibit lesser depletion or non-depletion. In contrast, Ti \citep[$T_{\rm cond,~Ti}=1582$ K; ][]{lodders2003CondensationTemperatures} is a refractory element expected to be significantly more depleted for post-RGB/post-AGB binary stars, causing [Zn/Ti] ratio to grow up to 3.4 dex \citep[as in case of AF Crt;][]{kluska2022GalacticBinaries}. We found that our targets exhibit a rather mild depletion with [Zn/Ti] values of 0.4 dex for SZ~Mon and 0.3 dex for DF~Cyg, as opposed to more extreme literature values of 0.8 dex and --0.7 dex, respectively.

The turn-off temperature $T_{\rm turn-off}$ \citep{oomen2019depletion} marks the onset of chemical depletion pattern, separating the relatively non-depleted elements (i.e., volatile elements like Zn and S) and the elements affected by the depletion process (i.e., refractory elements like Fe and Ti). It is worth noting that when comparing the elemental abundances of SZ~Mon and DF~Cyg to those of the post-AGB sample (roughly selected from \cite{oomen2019depletion} by their SED luminosity $L_{\rm SED}>2\,500~L_\odot$; see Section~\ref{ssec:obslum}), we notice that $T_{\rm turn-off}$ in SZ~Mon and DF~Cyg ($\approx1\,400$~K; see Fig.~\ref{fig:fnlabs}) is higher than the median $T_{\rm turn-off}$ in binary post-AGB sample ($\approx1\,100$~K; see \cite{oomen2019depletion} and references therein).

Regarding carbon isotopic ratios, our study marks the first attempt at estimating them in post-RGB stars. Using E-iSpec, as explained in Section~\ref{sssec:abuiso}, we derived $^{12}$C/$^{13}$C=8$\pm$4 for SZ~Mon and $^{12}$C/$^{13}$C=12$\pm$3 for DF~Cyg. To better understand the implications, we compared these values with theoretical predictions from the ATON evolutionary models, outlined in Section~\ref{sec:mod}.

\section{Integration of E-iSpec with ATON stellar evolutionary models}\label{sec:mod}
Despite the expected chemical depletion pattern observed in our targets (see Section~\ref{ssec:sanabs}), the derived CNO abundances deviate from both volatile elements (e.g., S and Zn) and refractory elements (e.g., Fe, Ti, etc.). In this section, we aim to ascertain whether combined knowledge of luminosity, surface CNO abundances, and carbon isotopic ratios could serve as valuable indicators of the nucleosynthetic history during the RGB phase before transitioning to the post-RGB phase.

\begin{figure*}
    \centering
    \includegraphics[width=.49\linewidth]{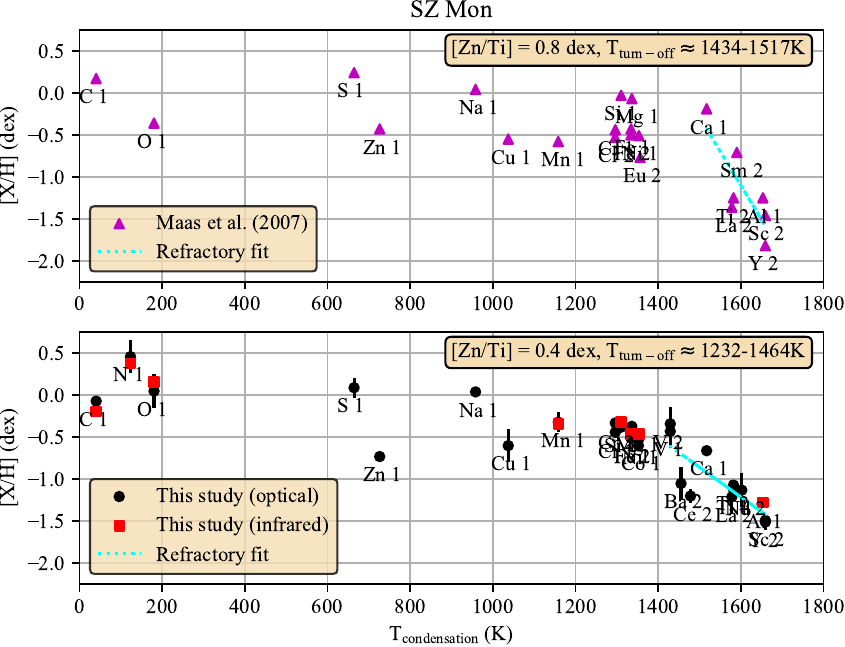}
    \includegraphics[width=.49\linewidth]{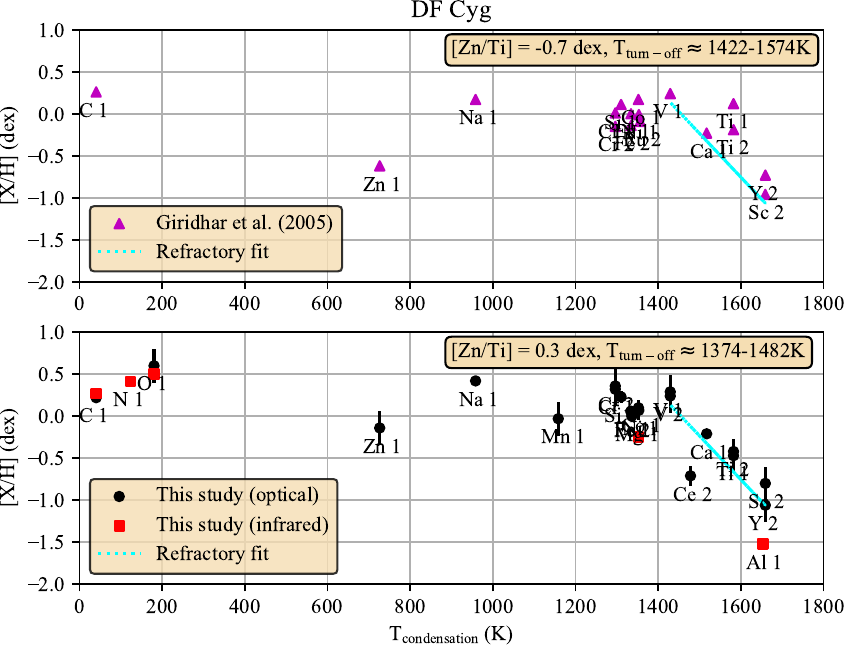}
    \caption{Abundances of SZ Mon (left) and DF Cyg (right) as functions of condensation temperature \citep{lodders2003CondensationTemperatures}. Top panels show the chemical analysis results from the literature \citep[][for SZ~Mon and DF~Cyg, respectively]{maas2007t2cep, giridhar2005rvtau}, bottom panels show the results from this study. Linear fits of the decrease in the abundances of the refractory elements are shown in cyan. The legend for the symbols and colours used are included within the plot.}\label{fig:fnlabs}
\end{figure*}

To do this, we compared the $^{12}$C/$^{13}$C ratio, [N/H], and C/O ratio of our two post-RGB targets (SZ~Mon and DF~Cyg, see Tables~\ref{tab:fnlabsSZM} and \ref{tab:fnlabsDFC}, and Fig.~\ref{fig:fnlabs}) presented above with results from stellar evolution modelling, obtained with the ATON code for stellar evolution \citep{ventura98}. We considered evolutionary sequences of low-mass stars, running from the pre-Main Sequence stage through the core H-burning and the ascending of the RGB, until the ignition of the helium flash. We note that for both targets we assumed the initial mass of $1~M_{\odot}$ based on the evolutionary sequences for post-AGB/post-RGB single stars presented in \cite{kamath2023models}.

While the ATON evolutionary models and the models from \cite{kamath2023models} are primarily designed for single stars, we used these models to reproduce the chemical composition of our post-RGB binary targets and therefore investigate the extent, to which binary interactions affect the stellar chemical composition. Specifically, we examined whether the chemical composition, especially the CNO abundances, observed in the post-RGB phase, could reflect the nucleosynthesis occurring during the RGB phase before this phase is terminated by the binary interaction.

We chose the luminosity as an indicator of the evolutionary stage, as use of time would be of little help in this case, given the significant difference among the duration of the various evolutionary phases. Given that the luminosities of SZ~Mon and DF~Cyg are significantly below $\rm 1.5\times10^3~L_{\odot}$ (see Section~\ref{ssec:obslum}), we safely assumed that these objects never entered the thermal-pulse phase \citep{ventura2022InternalProcesses}. However, the possibility that SZ~Mon and DF~Cyg are currently evolving through the early AGB phase following the exhaustion of central helium can not be ruled out completely. If we consider the evolutionary phases running from the tip of the RGB $L_{\rm RGB\ tip}\approx2\,500 L_\odot$ \citep{kamath2016postRGBs} until the first few thermal pulses, possible changes in the surface chemistry might be caused by the action of the second dredge-up in stars of initial mass above $\rm 3~M_{\odot}$, or by non-canonical deep mixing during the helium flash. The first hypothesis can be ruled out in the present context, as the luminosities at which the second dredge-up occurs are $\rm \sim 10^4~L_{\odot}$ or more, thus inconsistent with the values reported in Table~\ref{tab:varpro}. The possibility that SZ~Mon and DF~Cyg  are currently in a post-HB phase, and that their surface chemistry was altered by some deep mixing episode, such as those invoked by \citet{schwab2020HeFlashMixing} to explain the presence of lithium-rich giants, can not be excluded. However, we consider this option unlikely, as the deep mixing would favour lithium and $^{12}$C enrichment, whereas the valued reported in Tables~\ref{tab:fnlabsSZM} and \ref{tab:fnlabsDFC} indicate the typical effects of deep mixing of the surface convection with material processed by CNO cycling, typical of the RGB evolution. Therefore, we believe that the modelling of the RGB phase sufficiently characterises the two post-RGB targets.

The standard evolutionary sequences used in this study were calculated by using a diffusive approach to couple mixing of chemicals in regions unstable to convective motions and nuclear burning, following the classic scheme proposed by \citet{diffus}, where the diffusion coefficient $\rm D_{conv}$ is assumed to be proportional to convective velocities. To model the effects of deep extra mixing connected to the thermohaline instability, we also calculated evolutionary sequences where after the RGB bump the diffusion coefficient is estimated by adding to $\rm D_{conv}$ a term proportional to the gradient of the molecular weight \citep[$\rm D_t$; see Eq.~3 in][]{thermohaline}. As a first try, we used the same value of the proportional coefficient for the molecular weight \citep[$\rm C_t$; see Eq.~4 in][]{thermohaline}, namely $\rm C_t=1000$. To explore the sensitivity of the results obtained from this assumption, we also considered higher values of $\rm C_t=3000$, thus simulating the effects of very deep mixing experienced by the stars after the RGB bump.

\begin{table}
    \centering
    \caption{Final chemical analysis results for SZ Mon. Atmospheric parameters and elemental abundances were derived from SH\#73, unless stated otherwise. Chemical analysis results derived from other visits are specified in Table \ref{tabA:tststp}.}\label{tab:fnlabsSZM}
    \begin{tabular}{p{1cm}p{0.5cm}p{1.75cm}p{1.75cm}}\hline
        \multicolumn{4}{c}{SZ Mon} \\\hline
        \multicolumn{2}{c}{$T_{\rm eff}$ = 5460$\pm$60 K} & \multicolumn{2}{c}{$\log g$ = 0.93$\pm$0.10 dex} \\
        \multicolumn{2}{c}{[Fe/H] = --0.50$\pm$0.05 dex} & \multicolumn{2}{c}{$\xi_t$ = 4.37$\pm$0.08 km/s} \\\hline
        Element & \begin{tabular}{c} A09$^a$\\ $\log\varepsilon_\odot$ (dex) \end{tabular} & \begin{tabular}{c} This study \\ $[$X/H$]$ (dex) \end{tabular} & \begin{tabular}{c} M07$^b$ \\ $[$X/H$]$ (dex) \end{tabular} \\ \hline
        \ion{C}{i} & 8.43 & --0.07$\pm$0.05 & +0.17\\
        \ion{N}{i} & 7.83 & +0.38$\pm$0.06$^c$ & --\\
        \ion{O}{i} & 8.69 & +0.16$\pm$0.14$^c$ & --0.36\\
        \ion{Na}{i} & 6.24 & +0.04$\pm$0.05 & +0.04\\
        \ion{Mg}{i} & 7.60 & --0.37$\pm$0.04 & --0.07\\
        \ion{Al}{i} & 6.45 & --1.29$\pm$0.01$^c$ & --1.25\\
        \ion{Si}{i} & 7.51 & --0.38$\pm$0.05 & --0.03\\
        \ion{S}{i} & 7.12 & +0.09$\pm$0.12 & +0.24\\
        \ion{Ca}{i} & 6.34 & --0.66$\pm$0.03 & --0.19\\
        \ion{Sc}{ii} & 3.15 & --1.50$\pm$0.07 & --1.46\\
        \ion{Ti}{i} & 4.95 & --1.12$\pm$0.01 & --\\
        \ion{Ti}{ii} & 4.95 & --1.07$\pm$0.06 & --1.25\\
        \ion{V}{i} & 3.93 & --0.43$\pm$0.20 & --\\
        \ion{V}{ii} & 3.93 & --0.34$\pm$0.20 & --\\
        \ion{Cr}{i} & 5.64 & --0.45$\pm$0.02 & --0.44\\
        \ion{Cr}{ii} & 5.64 & --0.34$\pm$0.01 & --0.53\\
        \ion{Mn}{i} & 5.43 & --0.32$\pm$0.12 & --0.58\\
        \ion{Fe}{i} & 7.50 & --0.51$\pm$0.05 & --0.43\\
        \ion{Fe}{ii} & 7.50 & --0.51$\pm$0.02 & --0.50\\
        \ion{Co}{i} & 4.99 & --0.60$\pm$0.06 & --\\
        \ion{Ni}{i} & 6.22 & --0.52$\pm$0.08 & --0.51\\
        \ion{Cu}{i} & 4.19 & --0.60$\pm$0.20 & --0.55\\
        \ion{Zn}{i} & 4.56 & --0.73$\pm$0.04 & --0.43\\
        \ion{Y}{ii} & 2.21 & --1.51$\pm$0.09 & --1.82\\
        \ion{Ba}{ii} & 2.18 & --1.05$\pm$0.20 & --\\
        \ion{La}{ii} & 1.10 & --1.21$\pm$0.20 & --1.36\\
        \ion{Ce}{ii} & 1.58 & --1.20$\pm$0.08 & --\\
        \ion{Nd}{ii} & 1.42 & --1.13$\pm$0.20 & --\\
        \ion{Sm}{ii} & 1.01 & -- & --0.71\\
        \ion{Eu}{ii} & 0.52 & -- & --0.77\\\hline
        $^{12}$C/$^{13}$C & $\sim$89 & 8$\pm$4 & -- \\\hline
    \end{tabular}\\
    \textbf{Note:} The given value comes from: $^a$\cite{asplund2009}, $^b$\cite{maas2007t2cep}, $^c$SA\#2.
\end{table}

\begin{table}
    \centering
    \caption{Final chemical analysis results for DF Cyg. Atmospheric parameters and elemental abundances were derived from DH\#83, unless stated otherwise. Chemical analysis results derived from other visits are specified in Table \ref{tabA:tststp}.}\label{tab:fnlabsDFC}
    \begin{tabular}{p{1cm}p{0.5cm}p{1.75cm}p{1.75cm}}\hline
        \multicolumn{4}{c}{DF Cyg} \\\hline
        \multicolumn{2}{c}{$T_{\rm eff}$ = 5770$\pm$70 K} & \multicolumn{2}{c}{$\log g$ = 1.92$\pm$0.09 dex} \\
        \multicolumn{2}{c}{[Fe/H] = 0.05$\pm$0.05 dex} & \multicolumn{2}{c}{$\xi_t$ = 3.97$\pm$0.03 km/s} \\\hline
        Element & \begin{tabular}{c} A09$^a$\\ $\log\varepsilon_\odot$ (dex) \end{tabular} & \begin{tabular}{c} This study \\ $[$X/H$]$ (dex) \end{tabular} & \begin{tabular}{c} G05$^c$ \\ $[$X/H$]$ (dex) \end{tabular} \\ \hline
        \ion{C}{i} & 8.43 & +0.22$\pm$0.04 & +0.26\\
        \ion{N}{i} & 7.83 & +0.41$\pm$0.08$^c$ & --\\
        \ion{O}{i} & 8.69 & +0.60$\pm$0.20 & --\\
        \ion{Na}{i} & 6.24 & +0.42$\pm$0.02 & +0.17$\pm$0.18\\
        \ion{Mg}{i} & 7.60 & +0.00$\pm$0.06 & --\\
        \ion{Al}{i} & 6.45 & --1.52$\pm$0.10$^c$ & --\\
        \ion{Si}{i} & 7.51 & +0.23$\pm$0.07 & +0.11$\pm$0.06\\
        \ion{Ca}{i} & 6.34 & --0.21$\pm$0.02 & --0.23$\pm$0.12\\
        \ion{Sc}{ii} & 3.15 & --0.80$\pm$0.20 & --0.96$\pm$0.32\\
        \ion{Ti}{i} & 4.95 & --0.47$\pm$0.20 & +0.12$\pm$0.16\\
        \ion{Ti}{ii} & 4.95 & --0.42$\pm$0.01 & --0.19\\
        \ion{V}{i} & 3.93 & +0.29$\pm$0.20 & +0.24$\pm$0.19\\
        \ion{V}{ii} & 3.93 & +0.24$\pm$0.20 & --\\
        \ion{Cr}{i} & 5.64 & +0.32$\pm$0.20 & +0.01$\pm$0.13\\
        \ion{Cr}{ii} & 5.64 & +0.36$\pm$0.20$^d$ & --0.15$\pm$0.07\\
        \ion{Mn}{i} & 5.43 & --0.03$\pm$0.20$^d$ & --\\
        \ion{Fe}{i} & 7.50 & +0.06$\pm$0.05 & +0.03$\pm$0.17\\
        \ion{Fe}{ii} & 7.50 & +0.05$\pm$0.02 & --0.11$\pm$0.18\\
        \ion{Co}{i} & 4.99 & +0.10$\pm$0.06 & +0.17$\pm$0.06\\
        \ion{Ni}{i} & 6.22 & +0.07$\pm$0.12 & --0.01$\pm$0.08\\
        \ion{Zn}{i} & 4.56 & --0.14$\pm$0.20 & --0.62\\
        \ion{Y}{ii} & 2.21 & --1.06$\pm$0.20 & --0.73\\
        \ion{Ce}{ii} & 1.58 & --0.71$\pm$0.12 & --\\
        \ion{Eu}{ii} & 0.52 & -- & --0.09\\\hline
        $^{12}$C/$^{13}$C & $\sim$89 & 12$\pm$3 & -- \\\hline
    \end{tabular}\\
    %\raggedright
    \textbf{Note:} The given value comes from: $^a$\cite{asplund2009}, $^b$\cite{giridhar2005rvtau}, $^c$DA\#1, $^d$DH\#26.\\
\end{table}

\subsection{SZ Mon}\label{ssec:conszm}
\begin{figure*}
    \centering
    \includegraphics[width=.70\linewidth]{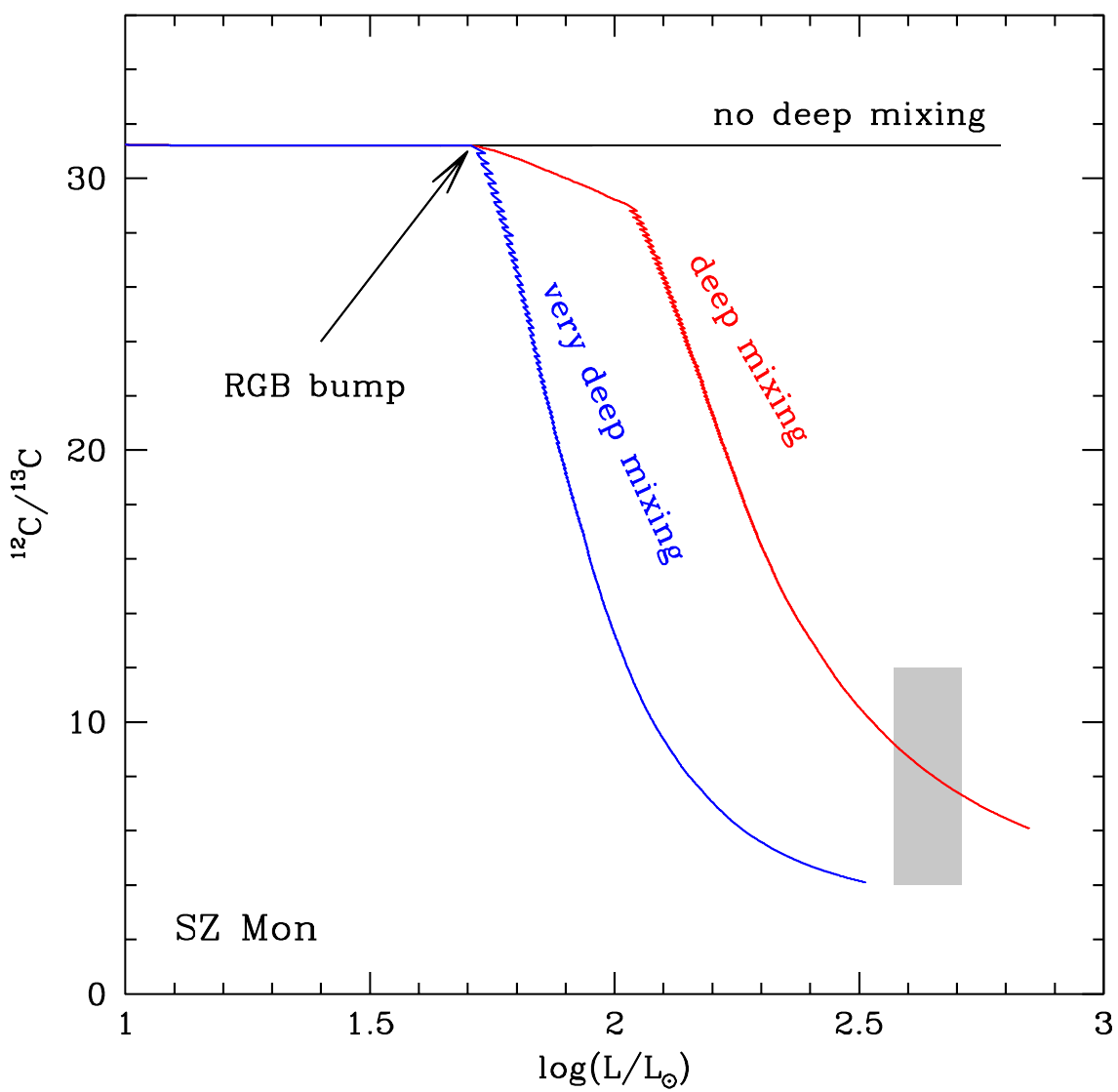}
    \includegraphics[width=.49\linewidth]{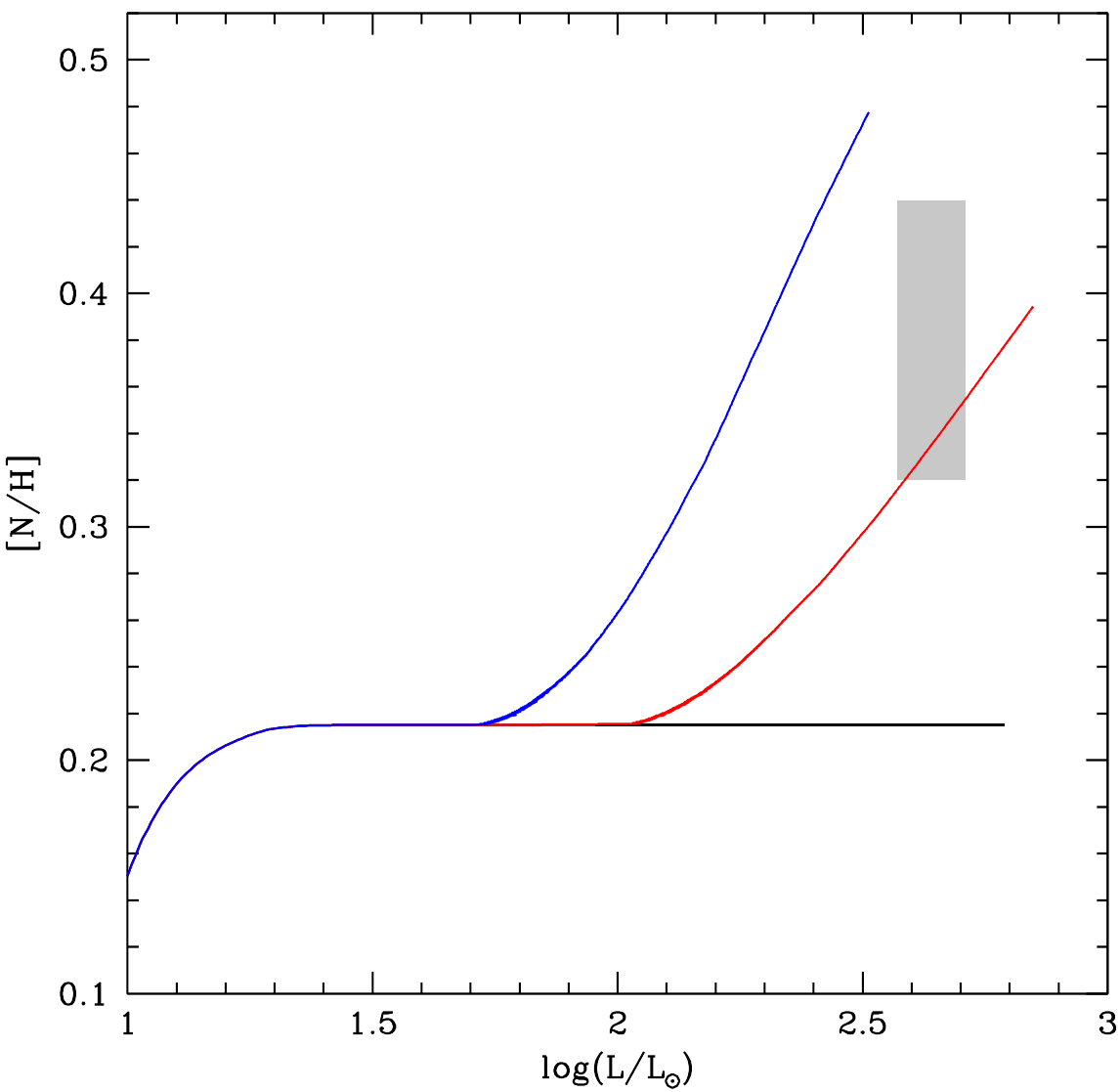}
    \includegraphics[width=.49\linewidth]{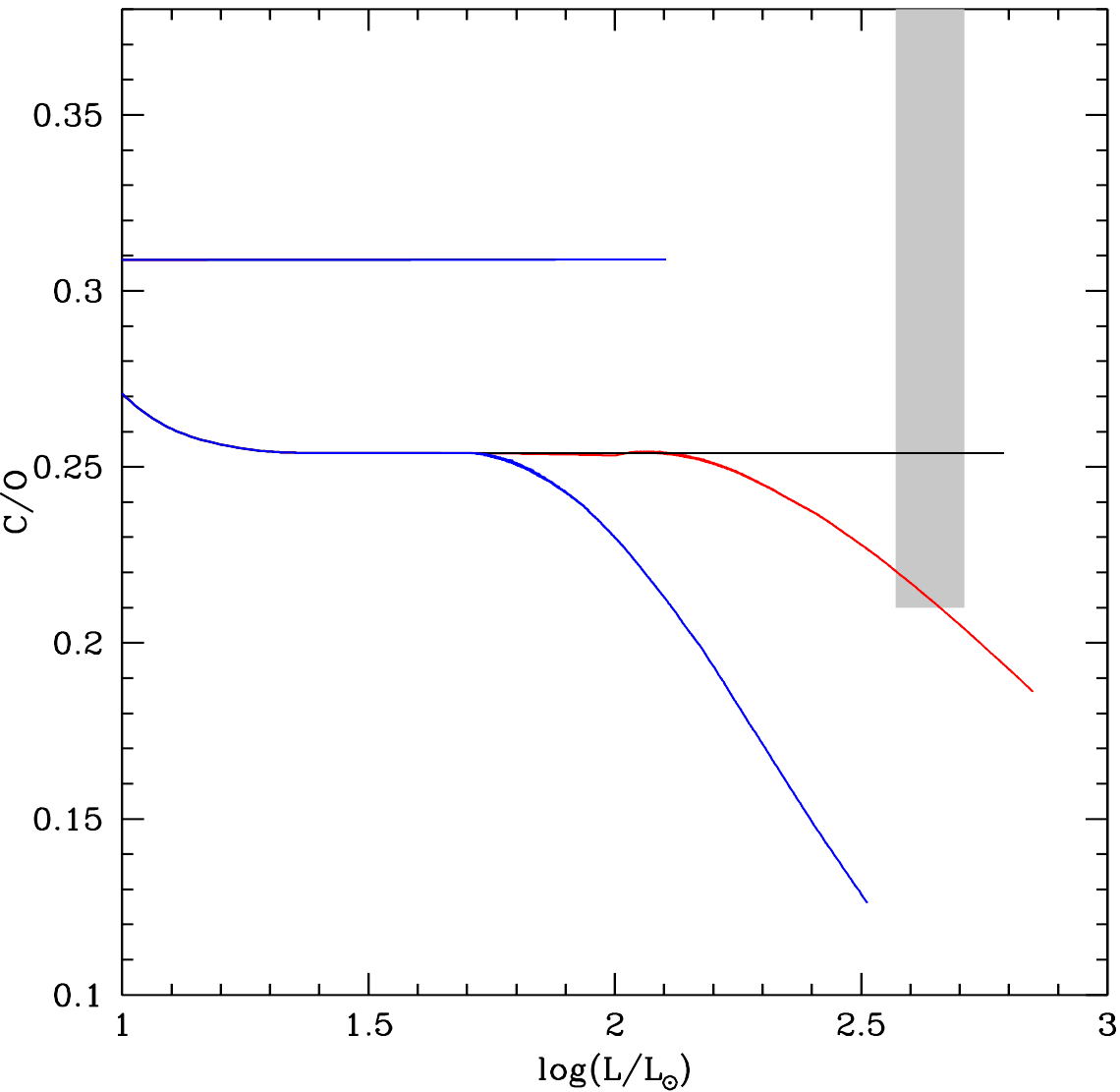}
    \caption{ATON evolutionary tracks of luminosity vs $^{12}$C/$^{13}$C (upper panel), luminosity vs [N/H] (lower left panel), and luminosity vs C/O (lower right panel) for a 1 M$_\odot$, Z=0.004 star. The solid lines are the evolutionary tracks of CNO abundances and carbon isotopic ratios for models assuming FDU and different levels of extra mixing: no deep mixing (black), moderately deep mixing (red), and very deep mixing (blue). We note that the flat line at C/O$\sim$0.31 (see lower right panel) represents the starting part (pre-MS) of the C/O evolutionary track. The grey boxes indicate the observed values for SZ~Mon within corresponding uncertainties (see Table~\ref{tab:fnlabsSZM}).}\label{fig:modszm}
\end{figure*}

To interpret the CNO abundances of SZ Mon and investigate its evolution, we modelled the evolution of $\rm 1~M_{\odot}$ star of metallicity $\rm Z=0.004$, which corresponded to initial metallicity of $\approx-0.7$ dex as traced by observed [S/H] and [Zn/H] (see Fig.~\ref{fig:fnlabs}). We note that the horizontal part of the evolutionary tracks shown in Fig.~\ref{fig:modszm} represents the initial pre-Main Sequence phase, during which the luminosity decreases and the surface chemistry is unchanged. During the RGB phase the luminosity increases, and the different sequences bifurcate after the RGB bump, when changes (due to extra mixing) in the surface chemical composition take place. In Fig.~\ref{fig:modszm}, we show the calculated models with no extra mixing (black lines), in which the only modification of the surface chemistry took place after the first dredge-up (FDU). We also assumed some further mixing after the bump, either moderate (red lines) or deep (blue lines). The grey boxes represent the corresponding observed values for SZ~Mon within their uncertainties.

The upper subplot of Fig.~\ref{fig:modszm} shows the evolution of the surface carbon isotopic ratio. It is clear that the assumption that no extra-mixing takes place after the RGB bump, thus the surface chemistry of the star is unchanged after the FDU (black line), is not consistent with the measured value (indicated with a grey box). The agreement of theoretical and experimental values of $^{12}$C/$^{13}$C becomes satisfactory, when moderately deep extra mixing is considered.

The lower left subplot of Fig.~\ref{fig:modszm} shows the evolution of the nitrogen abundance. We find that some extra mixing is required to fit the observations, as the FDU can account for a N enhancement of only $\sim0.2$ dex, against the $\sim0.4$ dex range derived from the spectroscopic analysis. Given that the efficiency of deep mixing affects the luminosity, at which a given surface nitrogen is reached, we deduce that use of a moderate deep mixing leads to a better agreement with the observations for [N/H].

Finally, in the lower right subplot of Fig.~\ref{fig:modszm}, we show the evolution of C/O ratio. The central value is more consistent with models without extra mixing, but the significant uncertainty in C/O of 0.11 (which is mostly caused by the uncertainty in oxygen abundance of 0.14 dex) does not allow us to draw any certain conclusions from this tracer.

Overall, the results presented so far indicate that SZ Mon evolved along the RGB until the core mass grew to $\rm \sim 0.36-0.37~M_{\odot}$, and the luminosity reached the nowadays value, in the $\rm 400-500~L_{\odot}$ range. At this evolutionary stage, mass transfer to the companion favoured the loss of the external mantle, an early departure from the RGB, and the beginning of the general contraction of the structure, until the current status was reached. The derived surface chemical composition, particularly the carbon isotopic ratio and nitrogen, suggests the combined effects of the FDU and deep mixing occurred after the RGB bump.

An alternative possibility is that the contraction to the blue started during the early AGB phase, after the exhaustion of central helium, before the occurrence of the first thermal pulse. Nevertheless, we contend that the evolution of both our targets was terminated during the RGB phase rather than the early-AGB phase. This is plausible, given the larger radius  of a star at the tip of the RGB compared to an early AGB star. Additionally, RGB timescales are about three times longer than early AGB timescales. Furthermore, if the star evolved until the tip of the RGB, the mixing mechanism would act for a longer duration, favouring a more pronounced modification of surface chemistry, such as a smaller carbon isotopic ratio and a higher abundance of nitrogen.

\subsection{DF Cyg}\label{ssec:condfc}
\begin{figure*}
    \centering
    \includegraphics[width=.70\linewidth]{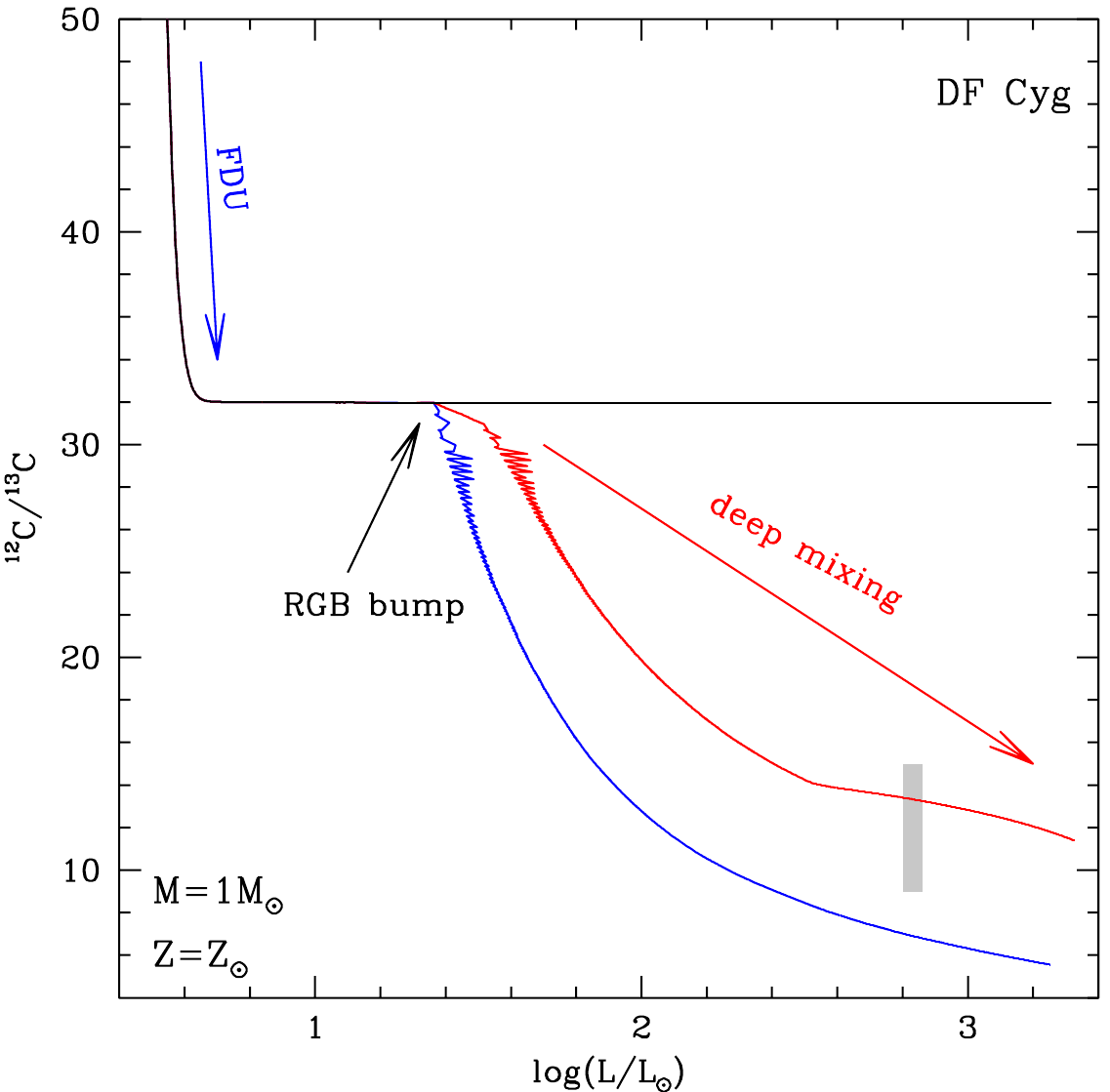}
    \includegraphics[width=.49\linewidth]{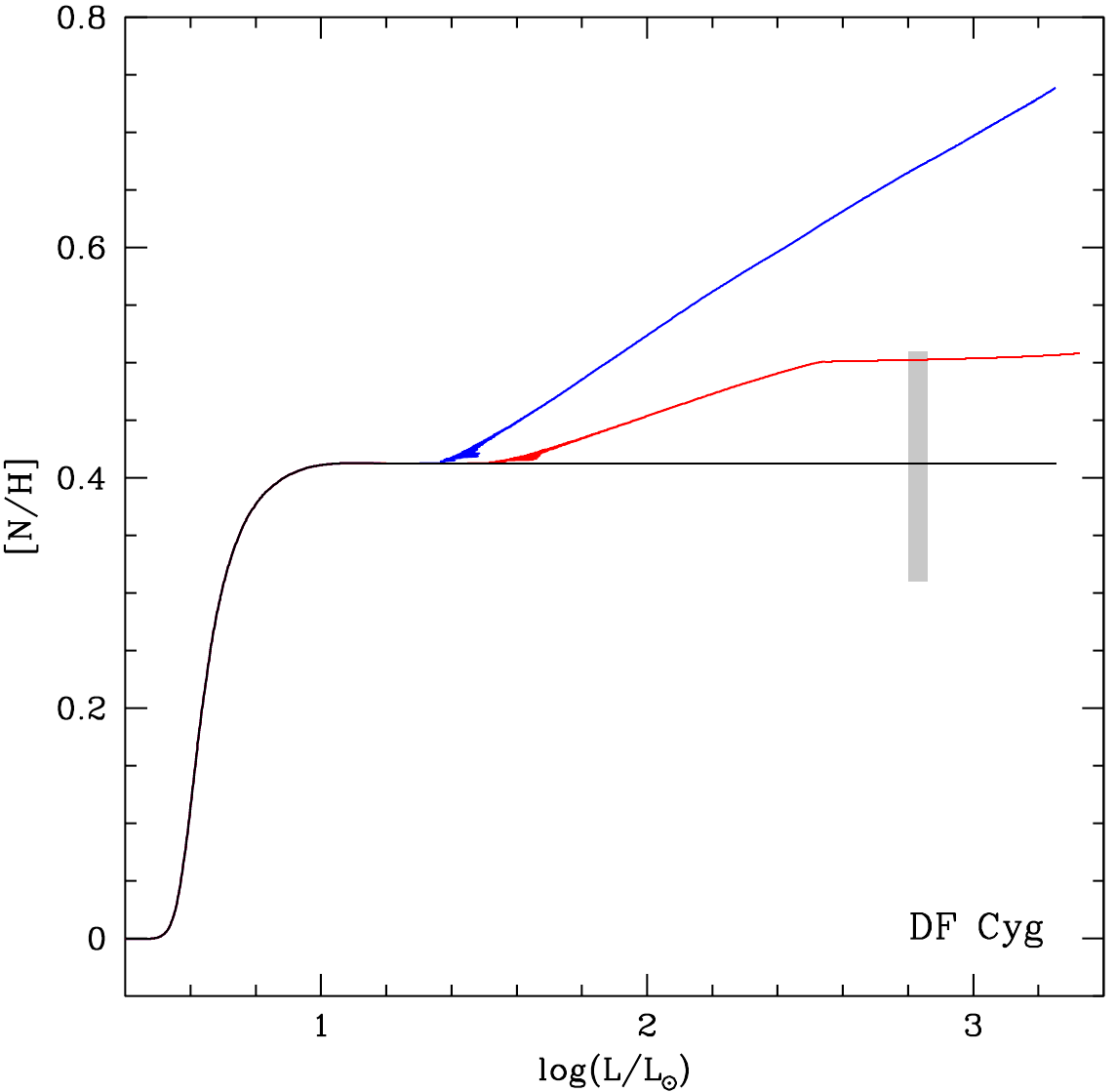}
    \includegraphics[width=.49\linewidth]{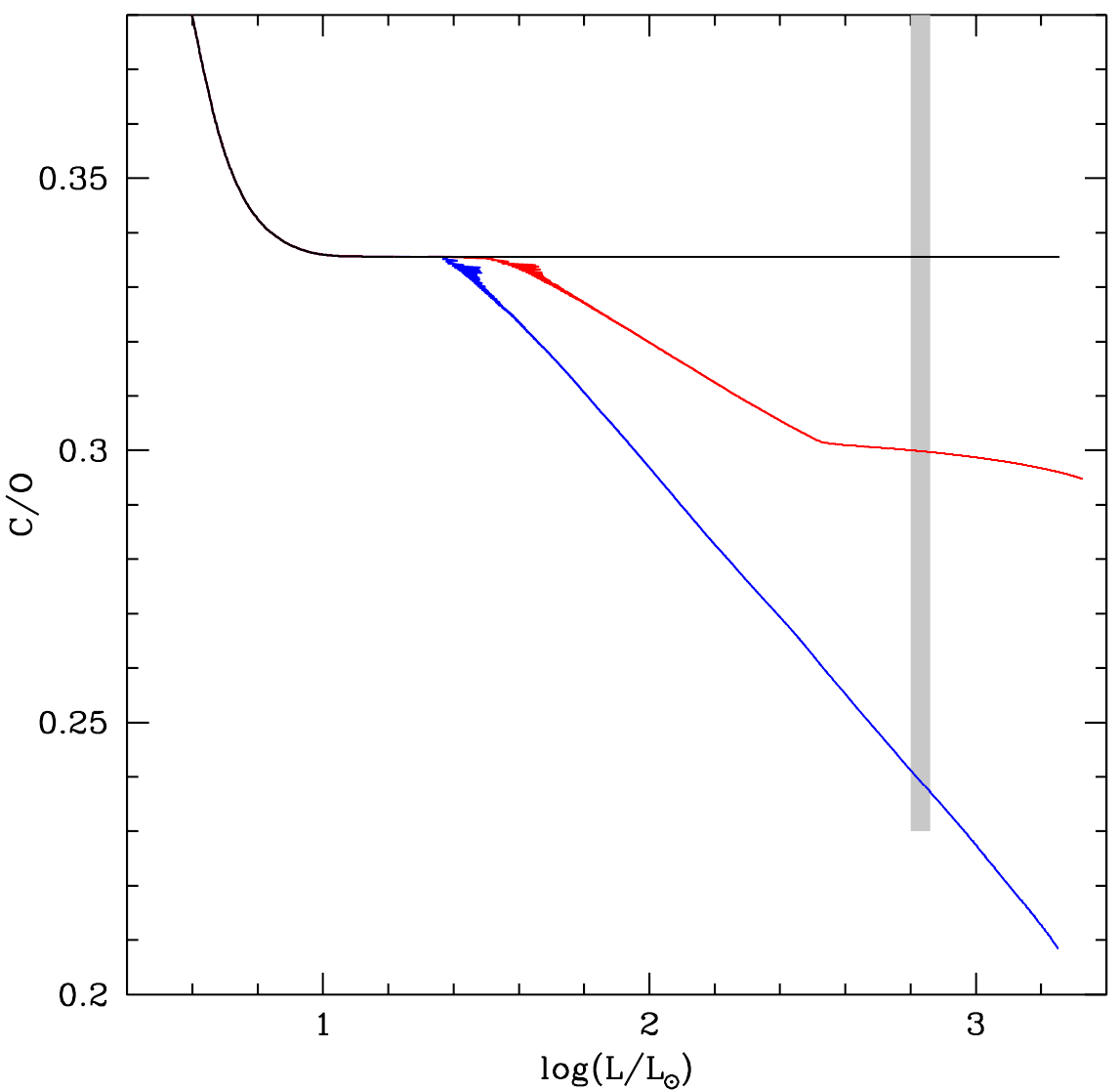}
    \caption{ATON evolutionary tracks of luminosity vs $^{12}$C/$^{13}$C (upper panel), luminosity vs [N/H] (lower left panel), luminosity vs C/O (lower right panel) for a 1 M$_\odot$, Z=Z$_\odot$ star. The solid lines are the evolutionary tracks of CNO abundances and carbon isotopic ratios for models assuming FDU and different levels of extra mixing: no deep mixing (black), moderately deep mixing (red), and very deep mixing (blue). The grey boxes indicate the observed values for DF~Cyg within corresponding uncertainties (see Table~\ref{tab:fnlabsDFC}).}\label{fig:moddfc}
\end{figure*}

To model the evolution of DF~Cyg, we constructed evolutionary sequences of a 1 $M_\odot$ star of solar metallicity. Similarly to SZ~Mon, the initial metallicity of DF~Cyg ($\approx0$ dex) was chosen according to [Zn/H] serving as a proxy (see Fig.~\ref{fig:fnlabs}). We note that the horizontal part of the evolutionary tracks shown in Fig.~\ref{fig:moddfc} represents the initial pre-Main Sequence phase, during which the luminosity decreases and the surface chemistry is unchanged. In Fig.~\ref{fig:moddfc}, the black lines indicate the results obtained by assuming only FDU, without any further extra mixing. The red and blue solid lines indicate the models involving extra mixing (moderately deep and very deep, respectively). The grey boxes in the different panels represent the observed range of the corresponding values within the errors.

The upper subplot of Fig.~\ref{fig:moddfc} shows the evolution of the carbon isotopic ratio. The observed value of DF~Cyg agrees well with the model predictions when some moderately deep extra mixing is considered (red line).

The lower left subplot of Fig.~\ref{fig:moddfc} shows the evolution of the nitrogen abundance. The substantial uncertainty of observed [N/H] for DF~Cyg (0.11 dex) prevents us from making definitive conclusions. However, we can deduce that experimental values of [N/H] agree with theoretical predictions from models involving either no extra mixing or moderately deep extra mixing.

In the lower right subplot of Fig.~\ref{fig:moddfc}, we show the evolution of the C/O with luminosity. Significant uncertainty in C/O (caused by uncertainty in [O/H] of 0.2 dex) does not allow us to use C/O as an indicator of extra mixing efficiency. However, for DF Cyg, the surface carbon was found to be super-solar (see Table~\ref{tab:fnlabsDFC} and Fig.~\ref{fig:fnlabs}). Since the surface carbon content is known to decrease during the RGB phase, we conclude that the matter, from which the star formed was slightly enhanced in carbon, with $\rm [C/Zn]_0\approx+0.3$. This value of initial carbon enhancement is within the scatter for Galactic solar type stars, as observed in the Galactic chemical history of carbon by \cite{nissen2020SolarTypeAbunds}.

We find an overall consistency between the results from stellar evolution modelling and the observations. The analysis of the behaviour of the surface carbon suggests that even a higher initial carbon would be consistent with the derived carbon abundance. However this would create some tension with the behaviour of nitrogen, which in that case would grow to values in excess of those consistent with the spectroscopic analysis.

Similarly to SZ~Mon, we argue that DF~Cyg underwent an early departure of the RGB, after the luminosity reached $\rm \sim 600-700~L_{\odot}$. In comparison with the evolution of SZ~Mon, we note that the effects of deep mixing are softer here, as clear in the slope of the $^{12}$C/$^{13}$C vs luminosity trends of the two stars. This is explained by the difference in the metallicity, as the temperature gradients in lower metallicity stars are higher, resulting in a more efficient transport process \citep{lagarde2012ExMixMod}.
%}


\section{Summary and Conclusions}\label{sec:con}
In this study, we used dusty post-RGB binary stars as tracers of the evolutionary processes and nucleosynthesis that occur during the RGB phase. To do this, we acquired multi-wavelength spectroscopic observations of SZ~Mon and DF~Cyg: high-resolution optical spectra obtained with the HERMES at Mercator telescope and mid-resolution NIR spectra from the APOGEE survey.

For our study, we developed E-iSpec, a semi-automatic Python-based tool, primarily to address the need of combining and analysing the HERMES (optical) and APOGEE (NIR) spectra. Existing optical spectral analysis tools (as well as APOGEE's internal pipeline ASPCAP) had significant limitations when dealing with spectra of evolved stars with complex atmospheres. E-iSpec, built upon iSpec (v2023.08.04), was specifically developed for spectral analyses of evolved stars. Updates included semi-automatic continuum normalisation, an updated molecular line list, error estimation for elemental abundances, and isotopic ratio calculations, substantially widened the usability of the tool. In this study, the atmospheric parameters and elemental abundances were derived from the atomic lines using EW method, while the carbon isotopic ratios were calculated from the molecular bands using SSF technique.

Using E-iSpec, we performed the chemical analysis of SZ~Mon and DF~Cyg (see Tables~\ref{tab:fnlabsSZM} and \ref{tab:fnlabsDFC}, respectively). We found that the depletion patterns of our post-RGB targets were similar to those of post-AGB binary stars from the sample of \cite{kluska2022GalacticBinaries}. However, for SZ~Mon and DF~Cyg, the chemical depletion pattern starts at higher condensation temperatures $T_{\rm turn-off,~post-RGB}\approx1\,400$~K than for the binary post-AGB sample ($T_{\rm turn-off,~post-AGB}\approx1\,100$~K). Even though our study is the first one to detect this prominent difference in $T_{\rm turn-off}$, a statistically comprehensive and methodologically homogeneous study with a larger sample of targets is required to validate whether the onset of the chemical depletion pattern occurs at higher condensation temperatures in post-RGB stars compared to their high-luminosity analogues, i.e, the post-AGB binary stars. This difference, if detected for the large samples, may hint at the different binary-disc interaction processes for post-AGB and post-RGB binary stars. However, this question is out of the scope of the current study and will be addressed in the future work.

Additionally, we derived the first carbon isotopic ratios in the atmospheres of the post-RGB stars ($^{12}$C/$^{13}$C$_{\rm SZ~Mon}=8\pm4$, $^{12}$C/$^{13}$C$_{\rm DF~Cyg}=12\pm3$). We investigated whether the binary interaction affected the CNO nucleosynthesis that occurred on the RGB and until the binary interaction prematurely terminated the RGB evolution of our targets. We investigated this by comparing the derived CNO abundances and the carbon isotopic ratios of our post-RGB binary targets with relevant theoretical yields from dedicated single star models calculated using the ATON stellar evolutionary code.

We found that CNO elemental abundances (i.e., [N/H] and C/O) and carbon isotopic ratios agreed well with the predictions from models, which experienced FDU with deep extra mixing. This agreement emphasised that in post-RGB binary targets, the observed CNO abundances reflect the chemical composition expected from single star nucleosynthesis (i.e., convective and non-convective mixing processes) occurring during the RGB phase before it is terminated by binary interaction. However, we stress that our findings are based on a limited sample size. In our ongoing study, our aim is to broaden our sample of targets, thereby facilitating a statistically comprehensive investigation, enabling us to validate and confirm the aforementioned results.

\section*{Acknowledgements}\label{sec:ack}
The spectroscopic results presented in this paper are based on observations made with the Mercator Telescope, operated on the island of La Palma by the Flemish Community, at the Spanish Observatorio del Roque de los Muchachos of the Instituto de Astrofisica de Canarias. This paper also includes data collected by the APOGEE survey.

MM1 acknowledges the International Macquarie Research Excellence Scholarship (iMQRES) program for the financial support during the research. MM1, DK, and MM2 acknowledge the ARC Centre of Excellence for All Sky Astrophysics in 3 Dimensions (ASTRO 3D), through project CE170100013. PV acknowledges the support received from the PRIN INAF 2019 grant ObFu 1.05.01.85.14 (“Building up the halo: chemo-dynamical tagging in the age of large surveys”, PI. S. Lucatello). HVW acknowledges support from the Research Council, KU Leuven under grant number C14/17/082. DAGH acknowledges the support from the State Research Agency (AEI) of the Spanish Ministry of Science and Innovation (MCIN) under grant PID2020-115758GB-I00/AEI/10.13039/501100011033. This article is based upon work from European Cooperation in Science and Technology (COST) Action NanoSpace, CA21126, supported by COST. TM acknowledges financial support from the Spanish Ministry of Science and Innovation (MICINN) through the Spanish State Research Agency, under the Severo Ochoa Program 2020-2023 (CEX2019-000920-S)

\section*{Data availability}
The data underlying this article are available in the article and in its online supplementary material.

\bibliographystyle{mnras}
\bibliography{BiblioList}
\label{lastpage}
%%%%%%%%%%%%%%%%%%%% APPENDICES %%%%%%%%%%%%%%%%%%
\appendix
\section{Evolved candidates observed with APOGEE}\label{app:tar}
In this Appendix, we present the initial target sample for this study, which consisted of 36 post-RGB/post-AGB candidates (see Table~\ref{tabA:allapo}). The MW (Milky Way, Galactic) targets were chosen from \cite{2011A&A...526A.162G, 2015MNRAS.453..133G, oomen2018OrbitalParameters, kamath2022GalacticSingles, kluska2022GalacticBinaries}, the SMC targets were selected from \cite{2014MNRAS.439.2211K}, and the LMC targets were chosen from \cite{2011A&A...530A..90V, 2011MNRAS.411.1597W, 2014MNRAS.439.1472M, 2015MNRAS.454.1468K}. The targets were selected by their IRAS, MSX and 2MASS colours such that they were feasible to be observed with the APOGEE survey (see Section~\ref{sssec:obsspcnir}) with an S/N greater than $\sim50$ needed for a precise chemical analysis. We filtered this initial sample using the following criteria:
\begin{enumerate}
    \item the final targets were previously observed with high-resolution optical spectrographs (UVES, HERMES). This allowed us to accurately determine atmospheric parameters and, if possible, the abundances of carbon, nitrogen, and oxygen (see Section~\ref{sec:san});
    \item the target's spectroscopically derived effective temperature should be below 5000 K, enabling the detection and study of CNO molecular bands (CO, CN).
\end{enumerate}

After performing the selection cuts, our final sample comprised two post-RGB binary stars: SZ~Mon and DF~Cyg (see Section~\ref{sec:tar}).

\begin{landscape}
\begin{table}
    \centering
    \caption{Initial target sample containing all confirmed post-AGB/post-RGB stars which were observed with APOGEE.}\label{tabA:allapo}
    \begin{tabular}{|l|c|c|c|c|c|c|c|}
    \hline
        Galaxy & 2MASS name & IRAS/SAGE name$^{a}$ & Other Names & R. A. & Dec. & SED Type & Reference \\
        ~ & ~ & ~ & ~ & (deg) & (deg) & ~ & ~ \\ \hline
        MW & J05075028+4824094 & IRAS 05040+4820 & BD+48 1220 & 076.959515 & 48.402634 & Shell & 1 \\
        ~ & J05365506+0854087 & IRAS 05341+0852 & -- & 084.229417 & 08.902422 & Shell & 1 \\
        ~ & J05405705+1014249 & IRAS 05381+1012 & BD+10 845 & 085.237725 & 10.240270 & Shell & 1 \\
        ~ & J06512784-0122158 & IRAS 06489-0118 & SZ Mon & 102.866018 & -01.371072 & Disc & 1 \\
        ~ & J06553181-0217283 & IRAS 06530-0213 & -- & 103.882582 & -02.291199 & Shell & 1 \\
        ~ & J15183614+0204162 & IRAS F15160+0215 & -- & 229.650616 & 02.071189 & Non-IR excess & 1 \\
        ~ & J15585827+2608046 & - & BD+26 2763 & 239.742810 & 26.134620 & Non-IR excess & 1 \\
        ~ & J17470327+2319454 & IRAS 17449+2320 & BD+23 3183 & 266.763660 & 23.329260 & Disc & 2 \\
        ~ & J19213906+3956080 & IRAS 19199+3950 & HP Lyr & 290.412790 & 39.935580 & Disc & 3 \\
        ~ & J19361378+0704184 & - & V870 Aql & 294.057430 & 07.071800 & Non-IR excess & 1 \\
        ~ & J19485394+4302145 & IRAS 19472+4254 & DF Cyg & 297.224760 & 43.037370 & Disc & 1 \\ \hline
        SMC & J00383006-7303340 & J003829.99-730334.1 & SSTISAGEMC J003830.04-730334.1 & 009.625263 & -73.059448 & Uncertain & 4 \\
        ~ & J00443128-7305496 & J004431.23-730549.3 & SSTISAGEMC J004431.30-730549.9 & 011.130125 & -73.097028 & Uncertain & 4 \\
        ~ & J00444111-7321361 & J004441.03-732136.0 & OGLE SMC-LPV-4910 & 011.171296 & -73.360039 & Shell & 4 \\
        ~ & J00445628-7322566 & J004456.21-732256.6 & SSTISAGEMC J004456.26-732256.8 & 011.234533 & -73.382408 & Uncertain & 4 \\
        ~ & J00494423-7252088 & J004944.15-725209.0 & SSTISAGEMC J004944.19-725208.9 & 012.434298 & -72.869118 & Uncertain & 4 \\
        ~ & J00510723-7341334 & J005107.19-734133.3 & SSTISAGEMC J005107.24-734133.3 & 012.780143 & -73.692612 & Disc & 4 \\
        ~ & J00522223-7335376 & J005222.19-733537.6 & SSTISAGEMC J005222.23-733537.6 & 013.092458 & -73.593778 & Uncertain & 4 \\
        ~ & J00530734-7344045 & J005307.35-734404.5 & SSTISAGEMC J005307.36-734404.5 & 013.280625 & -73.734583 & Uncertain & 4 \\
        ~ & J00551576-7125168 & J005515.71-712516.9 & SSTISAGEMC J005515.76-712516.9 & 013.815672 & -71.421349 & Uncertain & 4 \\
        ~ & J00590901-7106486 & J005908.99-710648.6 & SSTISAGEMC J005909.01-710648.5 & 014.787543 & -71.113503 & Disc & 4 \\ \hline
        LMC & J04524318-7047371 & J045243.16-704737.3 & SSTISAGEMC J045243.16-704737.3 & 073.179954 & -70.793655 & Disc & 5 \\
        ~ & J04562323-6927489 & J045623.21-692749.0 & SSTISAGEMC J045623.21-692749.0 & 074.096708 & -69.463611 & Disc & 5 \\
        ~ & J04565524-6827330 & J045655.23-682732.9 & SSTISAGEMC J045655.23-682732.9 & 074.230206 & -68.459175 & Disc & 5 \\
        ~ & J05022115-6913171 & J050221.17-691317.2 & SSTISAGEMC J050221.17-691317.2 & 075.588208 & -69.221444 & Shell & 5 \\
        ~ & J05145312-6917234 & J051453.10-691723.5 & SSTISAGEMC J051453.10-691723.5 & 078.721250 & -69.289861 & Uncertain & 5 \\
        ~ & J05172873-6942469 & J051728.71-694246.7 & - & 079.369745 & -69.713028 & Shell & 5 \\
        ~ & J05184886-7002469 & J051848.84-700247.0 & SSTISAGEMC J051848.84-700247.0 & 079.703619 & -70.046387 & Shell & 5 \\
        ~ & J05190686-6941539 & J051906.86-694153.9 & SSTISAGEMC J051906.86-694153.9 & 079.778608 & -69.698326 & Shell & 5 \\
        ~ & J05213559-6951572 & J052135.62-695157.1 & SSTISAGEMC J052135.62-695157.1 & 080.398297 & -69.865906 & Uncertain & 5 \\
        ~ & J05214797-7009569 & J052147.95-700957.0 & SSTISAGEMC J052147.95-700957.0 & 080.449915 & -70.165810 & Disc & 5 \\
        ~ & J05220425-6915206 & J052204.24-691520.7 & SSTISAGEMC J052204.24-691520.7 & 080.517725 & -69.255730 & Disc & 5 \\
        ~ & J05221852-6950134 & J052218.52-695013.3 & SSTISAGEMC J052218.52-695013.3 & 080.577187 & -69.837074 & Disc & 5 \\
        ~ & J05254820-6937002 & J052548.17-693700.1 & - & 081.450854 & -69.616737 & Uncertain & 5 \\
        ~ & J05345377-6908020 & J053453.75-690802.0 & SSTISAGEMC J053453.75-690802.0 & 083.724040 & -69.133910 & Disc & 6 \\
        ~ & J05455567-7057305 & J054555.68-705730.3 & SSTISAGEMC J054555.68-705730.3 & 086.481984 & -70.958481 & Disc & 5 \\ \hline
    \end{tabular}\\
    \textbf{Notes:} $^{a}$IRAS/SAGE names were adopted from \cite{2014MNRAS.439.2211K} for SMC targets and from \cite{2015MNRAS.454.1468K} for LMC targets. References for SED types: $^1$\cite{2015MNRAS.453..133G}, $^2$Van Winckel (private communication), $^3$\cite{oomen2018OrbitalParameters}, $^4$\cite{2014MNRAS.439.2211K}, $^5$\cite{2015MNRAS.454.1468K}, $^6$\cite{2011A&A...530A..90V}.
\end{table}
\end{landscape}

\section{Additional data for SZ Mon and DF Cyg}\label{app:add}
In this Appendix, we provide the information about all optical and NIR observations of SZ~Mon and DF~Cyg.

In Tables~\ref{tabA:szmvis} and \ref{tabA:dfcvis}, we present a comprehensive summary of the optical and NIR observations conducted on two target stars: SZ~Mon and DF~Cyg, respectively. Table~\ref{tab:obslog} contains observational visits, which we selected for our chemical analysis.

\begin{table}
    \centering
    \caption{All optical and near-infrared visits of SZ Mon. This table is published in its entirety in the electronic edition of the paper. A portion is shown here for guidance regarding its form and content.}\label{tabA:szmvis}
    \begin{tabular}{c@{\hspace{0.2cm}}c@{\hspace{0.2cm}}c@{\hspace{0.2cm}}c@{\hspace{0.2cm}}c@{\hspace{0.2cm}}} \hline
        Visit \# & Obs. ID & BJD & Exp. (s) & RV (km/s) \\ \hline
        \multicolumn{5}{c}{\textit{Mercator + HERMES}} \\ \hline
        SH\#1 & 00260298 & 2455160.7710 & 800 & --19.53$\pm$0.02 \\
        SH\#2 & 00272588 & 2455217.4926 & 1200 & 5.49$\pm$0.05 \\
        SH\#3 & 00313718 & 2455497.7400 & 850 & 50.83$\pm$0.30 \\
        SH\#4 & 00314466 & 2455507.6421 & 800 & 27.24$\pm$0.06 \\
        SH\#5 & 00326956 & 2455572.6357 & 1000 & 8.47$\pm$0.07 \\
        \multicolumn{5}{c}{\ldots} \\ \hline
    \end{tabular}
\end{table}

\begin{table}
    \centering
    \caption{The same as Table~\ref{tabA:szmvis}, but for DF Cyg.}\label{tabA:dfcvis}
    \begin{tabular}{c@{\hspace{0.2cm}}c@{\hspace{0.2cm}}c@{\hspace{0.2cm}}c@{\hspace{0.2cm}}c@{\hspace{0.2cm}}} \hline
        Visit \# & Obs. ID & BJD & Exp. (s) & RV (km/s) \\ \hline
        \multicolumn{5}{c}{\textit{Mercator + HERMES}} \\ \hline
        DH\#1 & 00239921 & 2455010.5832 & 3149 & 2.04$\pm$0.47 \\
        DH\#2 & 00240154 & 2455013.6518 & 1800 & 6.03$\pm$0.31 \\
        DH\#3 & 00240155 & 2455013.6733 & 1800 & 5.85$\pm$0.32 \\
        DH\#4 & 00240156 & 2455013.6948 & 1800 & 5.79$\pm$0.30 \\
        DH\#5 & 00307940 & 2455475.4697 & 1200 & --53.72$\pm$0.17 \\
        \multicolumn{5}{c}{\ldots} \\ \hline
    \end{tabular}
\end{table}

In Table~\ref{tabA:optlst}, we provide the optical line lists for SZ~Mon and DF~Cyg. In Tables~\ref{tabA:nirlstszm} and \ref{tabA:nirlstdfc} (for SZ~Mon and DF~Cyg, respectively), we specify the spectral windows, which were sensitive to different elements and show the spectral features in these regions.

\begin{table*}
    \centering
    \caption{Optical line list. This table is published in its entirety in the electronic edition of the paper. A portion is shown here for guidance regarding its form and content.}\label{tabA:optlst}
    \begin{tabular}{r@{\hspace{0.25cm}}c@{\hspace{0.25cm}}c@{\hspace{0.25cm}}c@{\hspace{0.25cm}}|c@{\hspace{0.25cm}}c@{\hspace{0.25cm}}c@{\hspace{0.25cm}}|c@{\hspace{0.25cm}}c@{\hspace{0.25cm}}c@{\hspace{0.25cm}}} \hline
        \multicolumn{4}{|c|}{Atomic data} & \multicolumn{3}{c}{$W_{\lambda, {\rm SZ\,Mon}}$ (m\AA)} & \multicolumn{3}{|c|}{$W_{\lambda, {\rm DF\,Cyg}}$ (m\AA)} \\
        Element & $\lambda$ (nm) & $\log gf$ & $\chi$ (eV) & SH\#73 & SH\#29 & SH\#47 & DH\#83 & DH\#26 & DH\#54 \\ \hline
        \ion{C}{i} & 493.2049 & -1.658 & 7.685 & 93.9 & 96.6 & - & - & - & - \\
        \ion{C}{i} & 538.0325 & -1.616 & 7.685 & 86.3 & 95.1 & - & 106.0 & 112.6 & - \\
        \ion{C}{i} & 658.7610 & -1.003 & 8.537 & - & - & - & 80.3 & 93.4 & - \\
        \ion{C}{i} & 805.8612 & -1.275 & 8.851 & - & - & - & 43.6 & 58.4 & - \\ \hline
        \ion{N}{i} & 868.3403 & 0.105 & 10.330 & 50.9 & - & - & - & - & - \\ \hline
        \ion{O}{i} & 557.7339 & -8.204 & 1.967 & - & - & - & 49.3 & 49.8 & - \\
        \ion{O}{i} & 636.3776 & -10.258 & 0.020 & 47.2 & 65.2 & 73.5 & 66.4 & 58.6 & 37.2 \\ \hline
        \multicolumn{10}{c}{\ldots} \\ \hline
    \end{tabular}
\end{table*}
\begin{table}
    \centering
    \caption{Near-infrared line list for SZ\,Mon. Iso code is isotopic code in the format ``$Z_10Z_2.A_10A_2$'', where $Z_1, Z_2$ are atomic numbers of the elements making up the molecule, and $A_1, A_2$ are the corresponding atomic masses. $W_{\lambda, {\rm theor}}$ are the equivalent widths calculated for metallicity-scaled solar abundances as a proxy of each line's impact on the overall profile. This table is published in its entirety in the electronic edition of the paper. A portion is shown here for guidance regarding its form and content.}\label{tabA:nirlstszm}
    \begin{tabular}{c@{\hspace{0.15cm}}c@{\hspace{0.15cm}}c@{\hspace{0.15cm}}c@{\hspace{0.15cm}}c@{\hspace{0.15cm}}} \hline
        \multicolumn{5}{c}{SZ Mon} \\
        $\lambda$ (nm) & Type & Species & Iso code & $W_{\lambda, {\rm theor}}$ (m\AA) \\ \hline
        \multicolumn{5}{c}{C} \\ \hline
        \multicolumn{5}{c}{\textit{1688.9100-1689.1666}} \\ \hline
        1688.8645 & Mol & CO & 608.012016 & 13.63 \\
        1688.8988 & Mol & CO & 608.013016 & 6.21 \\
        1688.9209 & Mol & OH & 108.001018 & 1.27 \\
        1688.9371 & Mol & CO & 608.012017 & 30.77 \\
        1688.9390 & Mol & CO & 608.012016 & 2.52 \\
        1688.9473 & Ato & \ion{Fe}{i} & 26 & 13.2 \\
        \multicolumn{5}{c}{\ldots} \\ \hline
    \end{tabular}
\end{table}

\begin{table}
    \centering
    \caption{Near-infrared line list for DF\,Cyg. Iso code is isotopic code in the format ``$Z_10Z_2.A_10A_2$'', where $Z_1, Z_2$ are atomic numbers of the elements making up the molecule, and $A_1, A_2$ are the corresponding atomic masses. $W_{\lambda, {\rm theor}}$ are the equivalent widths calculated for metallicity-scaled solar abundances as a proxy of each line's impact on the overall profile. This table is published in its entirety in the electronic edition of the paper. A portion is shown here for guidance regarding its form and content.}\label{tabA:nirlstdfc}
    \begin{tabular}{c@{\hspace{0.15cm}}c@{\hspace{0.15cm}}c@{\hspace{0.15cm}}c@{\hspace{0.15cm}}c@{\hspace{0.15cm}}} \hline
        \multicolumn{5}{c}{DF Cyg} \\
        $\lambda$ (nm) & Type & Species & Iso code & $W_{\lambda, {\rm theor}}$ (m\AA) \\ \hline
        \multicolumn{5}{c}{C} \\ \hline
        \multicolumn{5}{c}{\textit{1557.6983-1558.6885}} \\ \hline
        1557.7381 & Mol & CO & 608.012016 & 2.70 \\
        1557.7448 & Mol & CO & 608.012016 & 2.69 \\
        1557.7506 & Mol & CN & 607.012015 & 4.84 \\
        1557.7526 & Ato & \ion{Fe}{i} & 26 & 1.33 \\
        1557.7581 & Mol & CO & 608.012016 & 2.71 \\
        1557.7783 & Mol & CO & 608.012016 & 2.67 \\
        \multicolumn{5}{c}{\ldots} \\ \hline
    \end{tabular}
\end{table}

\section{Comprehensive details on epoch selection}\label{app:epo}
In this Appendix, we extensively explain how we selected the optical (HERMES) and NIR (APOGEE) spectral visits, which we used for the analysis of chemical composition of two post-RGB binary targets, SZ~Mon and DF~Cyg (see Section~\ref{sec:san}).

First, we selected the APOGEE visits where we were able to detect prominent molecular bands in the spectra. This resulted in choosing a single NIR visit for each target (where we detected the molecular features of CO and CN): SA\#2 for SZ~Mon and DA\#1 for DF~Cyg.

However, to calculate the isotopic ratios using the APOGEE spectra, accurately derived atmospheric parameters are required (a model atmosphere). To derive these parameters, the ionisation analyses should be conducted, but the APOGEE spectra lacked the essential combination of spectral lines across different ionisation levels (e.g., Fe I and Fe II). Therefore, we used the HERMES spectra to determine atmospheric parameters and elemental abundances through excitation and ionisation analyses of atomic line transitions (see Section~\ref{sssec:stepar}). Naturally, we selected the HERMES visits with pulsation phases, which roughly matched those of SA\#2 and DA\#1 (out of 88 and 83 optical visits for SZ~Mon and DF~Cyg, respectively). Among these matching optical visits, we identified those visits, which exhibited the highest S/N. Specifically, for SZ~Mon, the chosen optical visit was identified as SH\#47, while for DF~Cyg, it was DH\#54.

Unfortunately, due to prominent blending in the above mentioned optical visits, the number of atomic spectral features we could analyse was significantly limited. On the contrary, at hotter phases the number of spectral features was smaller, hence the level of blending in the spectra was decreased. So, we included two additional optical visits for each target with the highest S/N values, which occurred at pulsation phases with higher temperatures. This approach enabled us to confirm the atmospheric parameters derived from SH\#47 and DH\#54, as well as to extend the number of studied chemical species. These visits are provided in Table~\ref{tab:obslog}: SH\#73 and SH\#29 for SZ~Mon, DH\#83 and DH\#26 for DF~Cyg.

Using the selected visits, we derived the atmospheric parameters, the chemical abundances, and the carbon isotopic ratios for SZ~Mon and DF~Cyg (see Section~\ref{sec:san}).

\section{E-iSpec testing highlights}\label{app:tst}
In this Appendix, we provide the information about the most important tests we performed for validating performance of E-iSpec: 
\begin{enumerate}
    \item derivation of atmospheric parameters of a diverse sample of post-AGB stars using EW method,
    \item calculation of elemental abundances of SZ~Mon and DF~Cyg using EW method (for optical visits) and SSF technique (for NIR visits) in all observations, which were selected for chemical analysis (see Section~\ref{sec:obs}),
    \item SSF of atomic (Fe) and molecular (C, N, and O) features in giant branch stars.
\end{enumerate}

We note that we use EW method for a few reasons. Firstly, the synthetic spectral fitting technique involves an additional free parameter (the broadening velocity), which makes this technique more demanding in our case (for our calculation of the abundance uncertainties, we also need to account for the abundance deviation caused by the uncertainty of the broadening parameter). In contrast, the equivalent width method is insensitive to the broadening velocity. Since there is one less parameter in the procedure, this method is more robust. Secondly, in iSpec, the equivalent width method (Moog) provides the similar results as the synthetic spectral fitting technique (Turbospectrum), meaning that atmospheric parameters and elemental abundances obtained with these two methods match within the uncertainty ranges \citep{2014A&A...569A.111B}. This interrelation of Moog and Turbospectrum was tested for a large set of Gaia FGKM Benchmark Stars covering a wide range in $T_{\rm eff}$ (3500 to 6600 K), $\log g$ (0.50 to 4.60 dex) and [Fe/H] (--2.70 to 0.30 dex): the resulting median difference in $T_{\rm eff}$ between the two methods was found to be 50 K, the median difference in $\log g$ was --0.02 dex, the median difference in [Fe/H] was 0.03 dex, and the median difference in $\xi_{\rm t}$ was --0.34 km/s \citep{2019MNRAS.486.2075B}. Additionally, carbon isotopic ratios $^{12}$C/$^{13}$C, which are the main focus of this study, are independent of the elemental abundances of other elements (except for indirect impacts like blends or continuum placement), hence it is also independent of the method we use to derive the elemental abundances of other elements.

In Table~\ref{tabA:tstsmp}, we show the results of Fe lines analysis using E-iSpec. The testing sample consisted of previously studied post-AGB stars in the SMC and the LMC \citep{desmedt2012j004441, desmedt2015lmc2obj, kamath2017j005252, kamath2019depletionLMC}.

\begin{table*}
    \centering
    \caption{Testing EW method performance for chemically peculiar evolved stars. The atmospheric parameters are provided in the following format: literature/this study. The last line provides information on the ratio of automatically identified spectral lines out of those reported in the corresponding literature. The typical literature errors are: $\Delta T_{\rm eff}=250$ K, $\Delta\log g=0.5$ dex, $\Delta [$Fe/H$]$=0.5 dex, $\Delta\xi_{\rm t}=0.5$ km/s.}\label{tabA:tstsmp}
    \begin{tabular}{r@{\hspace{0.15cm}}c@{\hspace{0.25cm}}c@{\hspace{0.25cm}}c@{\hspace{0.25cm}}c@{\hspace{0.25cm}}c@{\hspace{0.25cm}}c@{\hspace{0.25cm}}} \hline
        Target & J051845$^a$ & J050356$^a$ & J050304$^a$ & J051848$^b$ & J004441$^b$ & J005252$^c$ \\ \hline
        $T_{\rm eff}$ & 5000/5080$\pm$80 & 5500/5600$\pm$80 & 5750/5740$\pm$30 & 6000/5880$\pm$100 & 6250/6290$\pm$100 & 8250/8260$\pm$340 \\
        $\log g$ & 0.50/0.39$\pm$0.18 & 1.00/1.20$\pm$0.15 & 0.00/0.04$\pm$0.05 & 0.50/0.49$\pm$0.21 & 0.50/0.63$\pm$0.17 & 1.50/1.08$\pm$0.06 \\
        $[$Fe/H$]$ & --1.10/--1.09$\pm$0.17 & --0.60/--0.54$\pm$0.06 & --2.60/--2.56$\pm$0.15 & --1.00/--1.21$\pm$0.19 & --1.34/--1.33$\pm$0.14 & --1.18/--1.03$\pm$0.14 \\
        $\xi_{\rm t}$ & 2.50/2.83$\pm$0.07 & 2.00/2.91$\pm$0.08 & 2.20/2.32$\pm$0.06 & 2.80/2.63$\pm$0.04 & 3.50/2.80$\pm$0.05 & 2.00/1.62$\pm$0.16 \\ \hline
        \begin{tabular}{r}
            Identified\\
            lines\\
            (ratio)
        \end{tabular} & \begin{tabular}{c}
            151/155 \\
            (97.42\%)
        \end{tabular} & \begin{tabular}{c}
            55/65 \\
            (84.62\%)
        \end{tabular} & \begin{tabular}{c}
            70/73 \\
            (95.89\%)
        \end{tabular} & \begin{tabular}{c}
            147/163 \\
            (90.18\%)
        \end{tabular} & \begin{tabular}{c}
            72/97 \\
            (74.23\%)
        \end{tabular} & \begin{tabular}{c}
            128/131 \\
            (97.71\%)
        \end{tabular} \\ \hline
    \end{tabular}\\
    \textbf{Notes:} The target is $^a$depleted in refractory elements, $^b$strongly enhanced in \textit{s}-process elements, $^c$depleted in \textit{s}-process elements.
\end{table*}

In Table~\ref{tabA:tstabs}, we provide the chemical analysis of two evolved stars from \cite{masseron2019APOGEE+BACCHUS} with the lowest $T_{\rm eff}$, hence with the most significant spectral blending. Since \cite{masseron2019APOGEE+BACCHUS} used the same NIR line list, we were able to independently confirm the performance of our SSF technique with evolved stars.

\begin{table*}
    \centering
    \caption{Testing SSF technique performance for evolved stars with similar line lists \citep[M+19 =][]{masseron2019APOGEE+BACCHUS}.}\label{tabA:tstabs}
    \begin{tabular}{l@{\hspace{0.25cm}}c@{\hspace{0.25cm}}c@{\hspace{0.25cm}}c@{\hspace{0.25cm}}c@{\hspace{0.25cm}}} \hline
        2MASS name & \multicolumn{2}{c}{J16323061-1306301} & \multicolumn{2}{c}{J19534103+1846056} \\
        Evol. stage & \multicolumn{2}{c}{RGB} & \multicolumn{2}{c}{eAGB} \\ \hline
        $T_{\rm eff}$ (K) & \multicolumn{2}{c}{4464} & \multicolumn{2}{c}{4745} \\
        $\log g$ (dex) & \multicolumn{2}{c}{1.46} & \multicolumn{2}{c}{1.86} \\
        $[$Fe/H$]$ (dex) & \multicolumn{2}{c}{--0.79} & \multicolumn{2}{c}{--0.54} \\ \hline
        \textbf{Source} & \textbf{M+19} & \textbf{This study} & \textbf{M+19} & \textbf{This study} \\
        A(Fe) (dex) & 6.71$\pm$0.03 & 6.66$\pm$0.18 & 6.96$\pm$0.20 & 7.00$\pm$0.14 \\
        A(C) (dex) & 7.54$\pm$0.10 & 7.55$\pm$0.12 & 7.53$\pm$0.18 & 7.51$\pm$0.11 \\
        A(N) (dex) & 7.56$\pm$0.04 & 7.71$\pm$0.15 & 8.44$\pm$0.09 & 8.60$\pm$0.06 \\
        A(O) (dex) & 8.74$\pm$0.04 & 8.66$\pm$0.14 & 8.93$\pm$0.11 & 8.86$\pm$0.14 \\ \hline
    \end{tabular}
\end{table*}

In Table~\ref{tabA:tststp}, we provide our results of chemical analysis performed on all selected observational visits of SZ~Mon and DF~Cyg to confirm the robustness of the final abundances. The atmospheric parameters of the infrared visits are bolded because they were fixed for these visits (see Section~\ref{ssec:saneis}).

\begin{table*}
    \centering
    \caption{Derived stellar parameters and abundances for optical and near-infrared visits of SZ Mon and DF Cyg. The abundances are provided in the form ``[X/H] (N)'', where N is number of spectral lines used to derive the abundance of the corresponding element.}\label{tabA:tststp}
    \begin{tabular}{rcccccccc}\hline
        & \multicolumn{4}{c}{SZ Mon} & \multicolumn{4}{c}{DF Cyg} \\
        Par$\backslash$Visit & 00866282 & 00397261 & 00458609 & APOGEE & 00972481 & 00412205 & 00574546 & APOGEE \\ \hline
        $T_\textrm{eff}$ & 5460$\pm$60 & 5420$\pm$80 & 4520$\pm$40 & \textbf{4500} & 5770$\pm$70 & 5750$\pm$70 & 5210$\pm$120 & \textbf{4500} \\
        $\log g$ & 0.93$\pm$0.10 & 0.78$\pm$0.09 & 0.94$\pm$0.14 & \textbf{1.00} & 1.92$\pm$0.09 & 1.71$\pm$0.09 & 2.13$\pm$0.26 & \textbf{2.00} \\
        $[$Fe/H$]$ & --0.50$\pm$0.05 & --0.50$\pm$0.10 & --0.52$\pm$0.10 & \textbf{--0.50} & 0.05$\pm$0.05 & --0.04$\pm$0.08 & 0.01$\pm$0.02 & \textbf{0.00} \\
        $\xi_\textrm{t}$ & 4.37$\pm$0.08 & 4.50$\pm$0.13 & 4.56$\pm$0.05 & \textbf{4.50} & 3.97$\pm$0.03 & 4.66$\pm$0.05 & 5.81$\pm$0.20 & \textbf{4.00} \\ \hline
        \ion{C}{i} & --0.07 (2) & 0.00 (2) & -- & --0.19 (1) & 0.22 (3) & 0.26 (3) & -- & 0.27 (3) \\
        \ion{N}{i} & 0.46 (1) & -- & -- & 0.38 (15) & -- & -- & -- & 0.41 (7) \\
        \ion{O}{i} & 0.05 (1) & 0.16 (1) & 0.07 (1) & 0.16 (2) & 0.60 (1) & 0.50 (1) & -- & 0.50 (2) \\
        \ion{Na}{i} & 0.04 (2) & 0.07 (2) & 0.12 (2) & -- & 0.42 (2) & 0.33 (2) & 0.35 (1) & -- \\
        \ion{Mg}{i} & --0.37 (2) & -- & --0.38 (3) & -- & 0.00 (2) & --0.08 (2) & -- & -- \\
        \ion{Al}{i} & -- & -- & -- & --1.29 (2) & -- & -- & -- & --1.53 (3) \\
        \ion{Si}{i} & --0.38 (6) & --0.35 (4) & --0.35 (4) & --0.38 (4) & 0.23 (5) & 0.24 (3) & 0.15 (1) & -- \\
        \ion{S}{i} & 0.09 (3) & 0.20 (2) & -- & -- & -- & -- & -- & -- \\
        \ion{Ca}{i} & --0.66 (8) & --0.73 (7) & --0.80 (2) & -- & --0.21 (3) & --0.28 (3) & --0.26 (3) & -- \\
        \ion{Sc}{ii} & --1.49 (4) & --1.58 (3) & -- & -- & --0.80 (1) & --0.83 (1) & --0.70 (1) & -- \\
        \ion{Ti}{i} & --1.12 (2) & --1.19 (2) & -- & -- & --0.47 (1) & -- & -- & -- \\
        \ion{Ti}{ii} & --1.07 (6) & --1.05 (6) & -- & -- & --0.42 (2) & --0.39 (2) & -- & -- \\
        \ion{V}{i} & --0.43 (1) & --0.24 (1) & --0.34 (6) & -- & 0.29 (1) & 0.24 (1) & 0.28 (5) & -- \\
        \ion{V}{ii} & --0.34 (1) & --0.33 (1) & -- & -- & 0.24 (1) & -- & -- & -- \\
        \ion{Cr}{i} & --0.44 (2) & --0.53 (1) & --0.47 (6) & -- & 0.32 (2) & 0.30 (1) & 0.41 (1) & -- \\
        \ion{Cr}{ii} & --0.33 (3) & --0.42 (3) & -- & -- & -- & 0.36 (1) & -- & -- \\
        \ion{Mn}{i} & --0.32 (3) & --0.37 (3) & --0.41 (2) & --0.34 (1) & -- & --0.03 (1) & 0.05 (1) & -- \\
        \ion{Fe}{i} & --0.50 (29) & --0.49 (21) & --0.54 (36) & --0.46 (38) & 0.06 (24) & --0.05 (24) & 0.04 (8) & -- \\
        \ion{Fe}{ii} & --0.51 (6) & --0.52 (5) & --0.52 (4) & -- & 0.05 (3) & --0.05 (3) & 0.01 (2) & -- \\
        \ion{Co}{i} & --0.60 (3) & --0.55 (2) & --0.56 (6) & -- & 0.10 (4) & 0.05 (3) & 0.06 (3) & --0.25 (1) \\
        \ion{Ni}{i} & --0.52 (13) & --0.51 (8) & --0.53 (3) & --0.43 (2) & 0.07 (17) & --0.02 (17) & --0.04 (3) & -- \\
        \ion{Cu}{i} & --0.60 (1) & --0.7 (1) & -- & -- & -- & -- & -- & -- \\
        \ion{Zn}{i} & --0.73 (2) & --0.66 (2) & -- & -- & --0.14 (1) & -- & -- & -- \\
        \ion{Y}{ii} & --1.51 (3) & --1.48 (3) & --1.53 (2) & -- & --1.06 (1) & --1.11 (1) & --0.83 (1) & -- \\
        \ion{Ba}{ii} & --1.05 (1) & --1.14 (1) & -- & -- & -- & -- & -- & -- \\
        \ion{La}{ii} & --1.21 (1) & -- & -- & -- & -- & -- & -- & -- \\
        \ion{Ce}{ii} & --1.20 (2) & --1.11 (2) & -- & -- & --0.71 (2) & --0.71 (2) & -- & -- \\
        \ion{Nd}{ii} & --1.13 (1) & --1.29 (1) & -- & -- & -- & -- & -- & -- \\ \hline
    \end{tabular}
\end{table*}

\bsp % typesetting comment
\end{document}